\documentclass[12pt,preprint]{aastex}
\slugcomment{DRAFT, 2001 Sep 10 [2]}
\begin{document}
\title{ENERGY BALANCE IN THE SOLAR TRANSITION REGION. IV.\\
       HYDROGEN AND HELIUM MASS FLOWS WITH DIFFUSION}
\author{{\sc J. M. Fontenla}\altaffilmark{1},
        {\sc E. H. Avrett}\altaffilmark{2}, and
        {\sc R. Loeser}\altaffilmark{2}}
\altaffiltext{1}{1874 MacCullen Drive, Erie, CO 80516,
       {\sf jfonten750@earthlink.net}}
\altaffiltext{2}{Smithsonian Astrophysical Observatory,
        Harvard-Smithsonian Center for Astrophysics,
        60 Garden Street, Cambridge, MA 02138,
       {\sf eavrett, rloeser@cfa.harvard.edu}}
%
%

\begin{abstract}

In this paper we have extended our previous modeling of energy balance in the
chromosphere-corona transition region to cases with particle and mass flows.
The cases considered here are quasi-steady, and satisfy the momentum and
energy balance equations in the transition region. We include in all
equations the flow velocity terms and neglect the partial
derivatives with respect to time. We present a complete
and physically consistent formulation and method for solving the non-LTE and
energy balance equations in these situations, including both particle diffusion
and flows of H and He.  Our results show quantitatively how mass flows
affect the ionization and radiative losses of H and He, thereby affecting
the structure and extent of the transition region.  Also, our computations
show that the H and He line profiles are greatly affected by flows.  We find that
line shifts are much less important than the changes in line intensity and central
reversal due to the effects of flows.  In this paper we use fixed conditions at the
base of the transition region and in the chromosphere because our intent is to show
the physical effects of flows and not to match any particular observations.
However, we note that the profiles we compute can explain the range of observed
high spectral and spatial resolution Lyman alpha profiles from the quiet Sun.
We suggest that dedicated modeling of specific sequences of observations based
on physically consistent methods like those presented here will substantially
improve our understanding of the energy balance in the chromosphere and corona.

\end{abstract}



\section{Introduction}

In our previous papers, Fontenla et al. (1990, 1991, 1993, hereafter FAL1,
FAL2, FAL3), we developed quasi-static models of the solar atmosphere, using
separate one-dimensional models to represent different quiet and active
solar features.  These models included a turbulent velocity, both to
broaden spectral lines and to add a Bernoulli term in the hydrostatic
equilibrium equation to represent a dynamic pressure contribution.
The modified equation was used to determine the density stratification
in the atmosphere that, because of this added term, departs from a
hydrostatic stratification that balances only gas pressure and gravity.
We introduced the important effects of hydrogen and helium diffusion in
the chromosphere-corona transition region, and for the first time obtained
reasonable agreement between calculated and observed line profiles for
hydrogen and helium, including general agreement with the well-observed
hydrogen Lyman alpha line profile (Fontenla, Reichmann, \& Tandberg-Hanssen
1988).

Many papers have studied quasi-steady flows in the transition region, e.g.,
Boris \& Mariska (1982), Craig \& McClymont (1986), McClymont \& Craig (1987),
Mariska (1988), and McClymont (1989). It is not possible for us to
address all of them; for a review, see Mariska (1992). Here we just mention
that these papers deal in detail with the upper transition region and coronal loops
where H and He are fully ionized, and use the radiative losses determined by Cox
\& Tucker (1969) for optically thin plasmas. These papers do not treat
accurately the lower transition region and chromosphere where H and He are
only partially ionized and where the resonance lines of H and He must be computed
from detailed solutions of the radiative transfer equations. Also, these papers
do not include the complicated processes of particle diffusion and radiative
transfer addressed by the FAL papers. Despite these shortcomings they indicated
that the observed redshifts in transition region lines might be explained by
quasi-steady flows in coronal loops.

Other, more recent calculations have been carried out, e.g., by Hansteen \& Leer
(1995), that include velocities in models assuming a fully ionized plasma.
While this assumption is valid in the corona, and perhaps (depending on the
velocity) in the upper transition region, it is inadequate for the low
transition region where H and He are only partially ionized.

The paper by Woods \& Holzer (1991) considers a multicomponent plasma composed of
electrons, protons, ionized helium, and minor ion species. These calculations
use the St. Maurice \& Schunk (1977a, b) treatment of particle diffusion and
heat flow (which is based on a different method but in most respects is equivalent
to that used in the FAL papers). Woods \& Holzer make the point that earlier
estimates of minor ion line intensities may not be accurate because 
the effects of flow and particle diffusion on these ion abundances
and ionization degrees were not included.

Hansteen, Leer \& Holzer (1997) carried out calculations that also include
ionization energy flow and particle diffusion, following again the St. Maurice
\& Schunk (1977a, b) formulation. They consider partial H and He ionization,
variable helium abundance, and possible differences between the electron,
H, and He temperatures, and they explore various possible parameters related
to the solar wind.  However, they do not include, consistently with particle
diffusion and flows, the detailed effects of H and He excitation, ionization,
and radiative losses.  Instead they make some simplifying assumptions such as:
1) taking photoionization rates from Vernazza, Avrett \& Loeser (1981)
that were determined for static empirical models without diffusion or energy
balance; 2) taking H, He, and other radiative losses from Rosner, Tucker
\& Vaiana (1978) (based on numerical expressions from the Cox \& Tucker (1969)
calculations) that do not include particle flows or diffusion and that are not
consistent with their own models.  As we showed in the FAL calculations,
these approximations differ very substantially from the values that result
from fully consistent solutions of the radiative transfer and statistical
equilibrium equations including particle diffusion. Moreover, as we show
here, particle flows also have important effects on the H and He line
intensities, radiative losses, and ionization rates (and consequently
on the ionization energy flux). These authors recognize that their
optically thin losses are not valid in the chromosphere
(we find that they are not valid in the lower transition region
either), and they handle these radiative losses in an ad hoc manner that we
believe is not very realistic because it does not include the correct
dependence of the radiative losses on the temperature and density
structure of the atmosphere. Despite these shortcomings, many important
conclusions result from their paper, although it is hard to discern how
some of them may be affected by the ad hoc assumptions made.

Chae, Yun, \& Poland (1997, hereafter CYP) have studied the effects of flow
velocities but they have not consistently solved all the statistical
equilibrium and radiative transfer equations. Instead, they assumed level
populations, ionization, and radiative losses that are inconsistent with the
velocities and the temperature stratifications that these authors use.
We comment further on the CYP results in \S 7.

Our work is new because we carry out consistent calculations of level populations
and ionization at all heights, and a consistent calculation of energy balance in
the lower transition region including the heights where H and He change from being
mostly neutral to almost fully ionized, and where the important optically thick
resonance lines of H and He are formed.  For completeness we also include energy
balance calculations in the upper transition region where H and He are almost
fully ionized, but here the results depend on approximate formulae for the
radiative losses due to other constituents.

Our previous calculations (FAL1, 2, 3) solved the H and He radiative transfer and
statistical equilibrium equations including particle diffusion, but assumed zero net
H and He particle fluxes and consequently zero mass flow. In the transition region
(except where strong inflows occur) diffusion is very important because the temperature
gradient, and thus the corresponding ionization gradient, is very large. Diffusion
causes a reduction of the ionization gradient, as a result of H atoms diffusing
outward and protons diffusing inward.  Helium diffusion is more complicated and shows
outward \ion{He}{1} and inward \ion{He}{3} diffusion, while \ion{He}{2} diffusion
varies with height. Also, the diffusion of hydrogen atoms and protons has a strong
effect on helium diffusion. In FAL2 and FAL3 we presented the fully consistent
formulation for static energy-balance cases, some numerical solutions
for typical portions of the solar atmosphere, and also some approximate formulas
that summarize these computations and that can be used to estimate radiative losses
and effective heat transport coefficients.

In the present paper we explore the effects of particle and mass flows,
in addition to diffusion, on the ionization and excitation of hydrogen and helium,
and on the energy balance. We have incorporated these physical processes into
the PANDORA computer program. Earlier versions of this program were used in the
FAL papers, and in previous ones discussed by Avrett \& Loeser (1992).
We again assume one-dimensional geometry, with height as the only measure of position.
We carry out our computations as in the FAL papers, but now we introduce prescribed
mass- and particle-conserving hydrogen and helium flows. Thus, we compute results for
two parameters, the flux $F_{\rm H}$ of hydrogen particles (H atoms and protons),
and the flux $F_{\rm He}$ of helium particles (\ion{He}{1}, \ion{He}{2}, and
\ion{He}{3}). We assume that these fluxes are constant through the transition region,
because if an initial change in a boundary condition occurs, the resulting dynamics
would rapidly lead to a flux that is constant with height in a time-scale comparable
with the sound travel-time across the region.  Since the lower transition region
(where 10$^4$ K $ < T < $ 10$^5$ K) is very thin (except, as we show, for strong
inflows), this travel time is a few seconds. Consequently, the constant
flux approximation is valid for cases where the velocity at the boundary
varies on time-scales of many seconds or longer (see our discussion in FAL3).  
In our calculation a time-dependent approach has no advantage since we are not dealing 
with explosive phenomena (such as the impulsive phase of solar flares) but rather with 
quiet and moderately active regions of the solar atmosphere that change in time-scales
of minutes.  As in our previous calculations we prescribe the temperature structure
of the underlying chromosphere and photosphere.  As the density rapidly increases with
diminishing height in the upper chromosphere, the flow velocities quickly become
very small but still have some effects in the upper chromosphere.  These effects are
fully included in our calculations since we include the velocity terms throughout
the chromosphere. The effects are negligible in the temperature minimum region and
the photosphere.  In this paper we present the method we use for these calculations,
and some of the results we obtained.

In our previous quasi-static models we determined the temperature vs. height
structure of the transition region by solving the energy balance equation.  In these 
models the radiative losses are balanced by the inward flux of heat (including
ionization energy) from the corona.  Part of the heat flux is due to thermal conduction, 
mainly by electrons (although H-atom conduction contributes at low ionization and low
temperature).  However, a large contribution to this inward energy flow is due to the 
ionization energy that the protons carry as they diffuse into the lower transition 
region.  Protons recombine at low temperatures and thus release their ionization energy.
Including such diffusion and inward energy flow leads to a smaller temperature gradient
and a more extended transition region than would result from thermal conduction alone.
In this paper we designate as heat flux the total of the flux of thermal, ionization,
and excitation energy, as well as the particle enthalpy flux.

Since we solve in detail only H and He, our models of the temperature variation
with height apply mainly to the low transition region, from the chromosphere to
about 10$^5$ K (depending on the hydrogen and helium flows) because at these temperatures 
H and He are the main contributors to the radiative losses and energy transport (in addition
to electron thermal conduction and enthalpy flow). At higher temperatures other species
dominate as H and He become completely ionized. More work is needed to carry out consistent 
calculations for species other than H and He, including both diffusion and flow velocity
(Fontenla and Rovira, in preparation). For the present we have approximated the effects
of these other species in the same way as in our previous papers by using the
Cox \& Tucker (1969) radiative losses, so that we can extend our calculations out
to coronal temperatures.

In this paper we present two sets of results. The first set shows the effects of
velocities on H and He ionizations for models that all have the same prescribed $T(z)$.
The second set shows the complete effects of velocities for models each having $T(z)$
in the transition region individually determined by energy balance.

The prescribed $T(z)$ in our first set of calculations is similar to the static
model C temperature distribution used by Fontenla et al. (1999), but slightly
modified at the base of the transition region, and extended to higher temperatures
in the low corona. Table 1 specifies this prescribed static model. The purpose of
these initial results is to show how velocities affect the ionization of H and He
in the transition region, without the additional complication of changing the
temperature structure. Thus, while the static model is in energy balance, the
models with flow, using the static temperature distribution, are not. We show the
equations and methods used to solve the statistical equilibrium and radiative
transfer equations including both diffusion and velocity terms. Also, we show the
ionization fractions and various line profiles for H and He calculated for
outward and inward mass flows compared with cases with the same flows but with
diffusion effects ignored. In this way we show that diffusion has significant
effects even in cases where the mass flow velocities are included.
   
After giving the results for the prescribed temperature distribution indicated above,
we compute the effects that mass flows have on the temperature stratification
of the transition region. We determine the temperature vs. height in these models
by including the heat conduction, enthalpy, and
ionization energy flux terms in the energy equation.
The radiative losses are computed from the detailed calculations described above,
and all computations are iterated until all the quantities are consistent with the
radiative transfer and statistical equilibrium equations as well as with the energy
balance and momentum balance equations.  The temperature structure of the
underlying chromosphere and photosphere must still be prescribed as indicated above
since we cannot compute the chromospheric temperature structure in energy balance.
That is because the mechanism of chromospheric heating is still unknown, and
because such a full energy balance computation requires knowledge of the dependence
of the heating on physical parameters such as height, density, electron density,
temperature, and magnetic field.

The purpose of our second set of calculations is to show how flow velocities affect
the equilibrium temperature vs. height structure of the transition region and how
this, in turn, affects the H and He lines.  We show the various temperature
distributions, the ionization fractions, radiative losses, and energy flux
variations with height and provide an interpretation of the results.  Finally,
we present the H and He line profiles for the resulting inflow and outflow models
and infer some relationships between various quantities that are useful
to interpret observations.


\section{Basic Equations}

Here we briefly review the basic theory for the various types of velocities, show how
these velocities are defined, and how they cause the ionization balance to differ from
the local static ionization balance.  Corresponding departures from local static
equilibrium can be expected in relative level populations as well, but in the cases
we consider these are less important than the ionization effects because of the
large transition rates between the levels of the atoms we study (a possible exception
is the lowest triplet energy level of the He atom).

The ionization equilibrium for ionization stage $i$ of a given element $k$ is described
by the equation
\begin{equation}
{ \partial n_{ik} \over \partial t } + \nabla \cdot
(n_{ik} {\bf V}_{ik}) = R_{ik} \, ,
\end{equation}
where $n_{ik}$ is the total number density of all energy levels in this ionization 
stage, $R_{ik}$ is the net creation rate for ions in stage $i$, and
${\bf V}_{ik}$ is the mean velocity of those ions.

Using $i+1$ and $i-1$ to refer to the next ionization stages higher and lower than $i$,
we can write
\begin{equation}
R_{ik} = n_{i+1,k} P_{i+1,i,k} + n_{i-1,k} P_{i-1,i,k} 
         - n_{i,k} (P_{i,i+1,k} + P_{i,i-1,k}) \, ,
\end{equation}
where $P_{i,j,k}$ is the transition rate to the final ionization stage $j$
per particle of the element $k$ in the initial ionization stage $i$.

The velocity ${\bf V}_{ik}$ can be decomposed into three components:
a) the center-of-mass velocity {\bf U} given by the motion of all types of
particles weighted by their mass and number density,
\begin{equation}
{\bf U} = { {\sum_{i,k} m_k n_{ik} {\bf V}_{ik} } \over {\sum_{i,k} m_k n_{ik} } } \, ,
\end{equation}
b) the velocity ${\bf v}_k$ of each element relative to the center-of-mass velocity,
\begin{equation}
{\bf v}_k = { {\sum_i n_{ik} {\bf V}_{ik} } \over {\sum_i n_{ik} } } - {\bf U}  \, ,
\end{equation}
and c) the diffusion velocity ${\bf v}_{ik}$ of each ionization stage relative to
$({\bf v}_k + {\bf U})$,
\begin{equation}
{\bf v}_{ik} = {\bf V}_{ik} - ({\bf v}_k + {\bf U}) \, .
\end{equation}
With these definitions, equation (1) may be written as
\begin{equation}
{ \partial n_{ik} \over \partial t } + \nabla \cdot
[ n_{ik} ( {\bf v}_{ik} + {\bf v}_k + {\bf U} ) ]
= R_{ik} \, ,
\end{equation}
showing how the three velocities are involved in determining $n_{ik}$. The
decomposition is useful because, as we explain below, these three components of
the velocity result from different physical phenomena, and different particle
and momentum conservation constraints apply to each component.

The lower transition region between the solar chromosphere and corona is
a very thin layer (at least in the static case) that occurs at various heights and
orientations above the photosphere. For our purposes this layer can be locally
approximated by a one-dimensional stratification. Thus, in equation (6) we
consider spatial variations as a function of only the height coordinate $z$.

Furthermore, the steady-state approximation is reasonable in the transition region
because ions are quickly transported, in a few seconds, by flows and diffusion
to the locations where they ionize or recombine. Thus, we drop the partial time
derivative term and write equation (6) as
\begin{equation}
{ d \over dz } [ n_{ik} ( v_{ik} + v_k + U ) ] = R_{ik} \, ,
\end{equation}
where $n_{ik} ( v_{ik} + v_k + U )$ is the ionization stage flow.

The mass velocity, $U$, is not affected by atomic collisions but only by macroscopic
forces, and the conservation of mass gives
\begin{equation}
{ d \over dz } ( \rho U ) = 0 \, ,
\end{equation}
where $\rho$ is the mass density. The integration of this equation defines the
constant mass flow, $F_m = \rho U$.

The velocity $v_k$ of a given element relative to $U$ is determined by the abundance
gradient and by various forces acting on the given element, moderated by elastic and
inelastic collisions between different elements and with free electrons.
In the relatively high densities of the solar atmosphere out to the lower corona
these velocities are expected to be in the diffusion regime where collisions
drive the particle distributions close to a Maxwellian function centered
around the velocity $U$.  Since we do not consider nuclear reactions,
the total number density
\begin{equation}
n_k = \sum_i n_{ik}
\end{equation}
of element $k$ satisfies the element conservation equation
\begin{equation}
{ d \over dz } [ n_k ( v_k + U ) ] = 0 \, .
\end{equation}
The integration of this equation defines the constant element flow,
\begin{equation}
F_k = n_k ( v_k + U ) \, .
\end{equation}

The remaining flow velocity $v_{ik}$ is that of ionization stage $i$
relative to the mean velocity $(v_k + U)$ of the element. This flow is
driven by various forces (including electric fields and ionization gradients),
and is slowed by collisions. Radiative and collisional interactions
transform ions from one stage to another, and in the transition region this
induces strong ionization gradients that lead to significant ion
diffusion velocities $v_{ik}$. Thus, the ionization stage flow in
equation (7) is not constant.

Using the element flow, $F_k$, we can write equation (7) as
\begin{equation}
{ d \over dz } [ n_{ik} ( v_{ik} + { F_k \over n_k }) ] = R_{ik} \, .
\end{equation}
Defining the ionization fraction $y_{ik} = n_{ik} / n_k$, we can write
\begin{equation}
{ d \over dz } [ y_{ik} ( n_k v_{ik} + F_k ) ] + r_{ik} y_{ik} = s_{ik}
\end{equation}
where, from equation (2),
\begin{eqnarray}
r_{ik} &=& n_k ( P_{i,i+1,k} + P_{i,i-1,k} ) \, , \nonumber \\
s_{ik} &=& n_{i+1} P_{i+1,i,k} + n_{i-1} P_{i-1,i,k} \, .
\end{eqnarray}
The system of equations (13) for all stages of ionization of a given element
is redundant, since the sum of all such
equations is zero. Thus, we consider these equations for all but the fully
ionized stage, and supplement these equations with equation (9).
In the next sections we use equation (9) to modify the above expressions
for $r_{ik}$ and $s_{ik}$ in order to obtain a set of
equations that is better conditioned than if equation (9) were used separately.

We can further transform equation (13) by changing the independent variable $z$
to $\zeta_{ik}$ where $d\zeta_{ik} = r_{ik} dz$ and letting
$\sigma_{ik} = s_{ik} / r_{ik}$, so that
\begin{equation}
{ d \over d \zeta_{ik} } [ y_{ik} ( n_k v_{ik} + F_k ) ] + y_{ik} = \sigma_{ik} \, .
\end{equation}
This shows that $\sigma_{ik}$ is the value $y_{ik}$ would have if all the
velocities were zero, i.e. in local static equilibrium. The transformation from
$z$ to $\zeta_{ik}$ is always possible because $r_{ik}$ is always larger than zero,
and $\zeta_{ik}$ is given by
\begin{equation}
\zeta_{ik}(z) = \int_{z_0}^z r_{ik}(z^\prime) dz^\prime
\end{equation}
where $z_0$ is the height of one of the boundaries of the region where equation
(13) is evaluated.


\section{Flow Effects}

In our previous papers we considered only diffusion, and assumed zero particle
flow for element $k$ so that $F_k = 0$.  We now first consider solutions
corresponding to non-zero particle flows without ionization-stage diffusion,
to illustrate how a formal solution of equation (15) can be obtained in this case.
These solutions are not physically meaningful but are presented only to gain
insight in the ways in which specified particle flows affect H and He ionization,
without the additional complication of ionization-stage diffusion.  Thus in this
section we consider $F_k \neq 0$ and $v_{ik} = 0$. Results with both particle
flow and diffusion are given in \S 4.

When $v_{ik} = 0$, equation (15) reduces to
\begin{equation}
F_k { d y_{ik} \over d \zeta_{ik} } + y_{ik} = \sigma_{ik}  \, ,
\end{equation}
which can be simply integrated from a boundary condition. In the
following we will omit all indices for simplicity. The solution is
well behaved when the boundary condition in equation (16) is chosen to be
closest to the origin of the flow, i.e., $z_0$ is the innermost height for an
outward flow, or the outermost height for an inward flow. The analytic solution is
\begin{equation}
y(\zeta) = y(0) \exp^{-\zeta/F} + \int_0^\zeta \sigma( \zeta^\prime )
     \exp^{-(\zeta - \zeta^\prime )/F} { d \zeta^\prime / F } \, .
\end{equation}
A reasonable value to impose at the boundary is to let $y(z_0) = y(\zeta = 0) =
\sigma (\zeta = 0)$. Note that $y$ approaches $\sigma$ as $F$ approaches zero.

This analytic solution is possible only when $v_{ik} = 0$ because in general
$v_{ik}$ cannot be specified {\it a priori} but has to be solved with the
diffusion theory that specifies its dependence on the density gradients, and
therefore on $n_{ik}$. The solutions for $v_{ik} \neq 0$ are
discussed later in this paper.


\subsection{Hydrogen Flow}

In the case of hydrogen we consider the proton number density $n_{\rm p}$ (the
\ion{H}{2} density), the atomic hydrogen number density $n_{\rm a} = \sum_\ell n_\ell$
(the \ion{H}{1} density), where the sum is over all bound levels $\ell$,
and the total hydrogen density $n_{\rm H} = n_{\rm a} + n_{\rm p}$.
Equation (2) for the net creation rate of neutral hydrogen atoms can be written as
\begin{equation}
R_{\rm a} = n_{\rm p} \sum_\ell P_{\kappa\ell} - \sum_\ell n_\ell P_{\ell\kappa}
          = ( n_{\rm H} - n_{\rm a} ) \sum_\ell P_{\kappa \ell} 
            - n_{\rm a} \sum_\ell \left( { n_\ell \over n_{\rm a} } \right)
            P_{\ell\kappa} \, ,
\end{equation}
where the index $\kappa$ designates the continuum (the ionized H, or proton, state),
and $P_{\kappa\ell}$ and $P_{\ell\kappa}$ are the recombination and ionization rates
to and from level $\ell$. The second form of the right hand side of the equation
illustrates how we handle the redundancy of the equations (7) for neutral and
ionized hydrogen: we use only the equation for neutral hydrogen and eliminate the
number density of ionized hydrogen using equation (9) for $n_{\rm H}$.

For hydrogen, equation (13) applies with $y_{\rm a} = n_{\rm a} / n_{\rm H}$, the
atomic hydrogen fraction, and $F_k = F_{\rm H}$ the total hydrogen particle flow,
but using equation (19) we replace equations (14) for $r$ and $s$ by
\begin{eqnarray}
r &=& n_{\rm H} \sum_\ell [ P_{\kappa \ell} + \left( { n_\ell \over n_{\rm a} }
         \right) P_{\ell \kappa} ] \nonumber \\
s &=& n_{\rm H} \sum_\ell P_{\kappa \ell} \, .
\end{eqnarray}

Using the above definition of the parameters $r$ and $s$, we obtain
\begin{equation}
\sigma = { {\sum_\ell P_{\kappa \ell}} \over  {\sum_\ell [ P_{\kappa \ell} + 
          ( { n_\ell \over n_{\rm a} } ) P_{\ell \kappa} ]} }
\end{equation}
for the term on the right side of equation (17).  The quantity $\sigma$ is the
value that $y_{\rm a}$ would have in the zero velocity case (with zero diffusion).

Due to radiative transfer, the quantities $r$ and $\sigma$ depend not only on
the local values of the ionization and recombination rates but also on the
level populations and ionization rates at other heights.
We start the calculations with values of $n_\ell / n_{\rm a}$
from previous calculations. Then, after every calculation of
$n_{\rm a}$ by the method described above, each level number density is
corrected by first normalizing it to the new $n_{\rm a}$ and then by computing the
solution of the statistical equilibrium and radiative transfer equations
for all levels, as is explained below.  This procedure is iterated until
convergence to the proper solution is achieved.

The net rate into bound level $m$ from the ionized stage and from all other bound
levels is given by
\begin{equation}
R_m = \sum_{\ell \neq m} n_\ell P_{\ell m} + n_\kappa P_{\kappa m}
      - n_m ( \sum_{\ell \neq m} P_{m \ell} + P_{m \kappa} ) \, ,
\end{equation}
and $R_{\rm a}$ in equation (19) is the sum of $R_m$ over all $m$ (since the
bound-bound transition terms cancel in this sum). An equation similar to
equation (12) can be written for each bound level $m$.
This equation in the case $v_{ik} = 0$ is
\begin{equation}
F_{\rm H} { d \over dz } \left( { n_m \over n_{\rm H} } \right) = R_m \, .
\end{equation}
which is the statistical equilibrium equation for the case of flows when
diffusion is ignored. Thus, to solve for the hydrogen level populations 
we use the equation
\begin{equation}
n_m ( \sum_{\ell \neq m} P_{m \ell} + P_{m \kappa} + G_m )
    = \sum_{\ell \neq m} n_\ell P_{\ell m} + n_\kappa P_{\kappa m} 
\end{equation}
where
\begin{equation}
G_m = { F_{\rm H} \over n_m } { d \over dz } \left( { n_m \over n_{\rm H}
        } \right) \, .
\end{equation}
We use the new $y_{\rm a}$ obtained above along with the
previous ratios $n_m / n_{\rm a}$ to obtain $n_m / n_{\rm H}$ and thus $G_m$.
Then we solve the set of statistical equilibrium equations (24) for all levels,
coupled with the radiative transfer equations, to obtain new values of $n_\ell$
and $n_{\rm p}$ with the constraint that they add up to the given $n_{\rm H}$.
Note that this solution involves excitation, de-excitation, ionization,
and recombination rates which depend on radiation intensities that in turn
depend on the number densities throughout the atmosphere according to the
equations of radiative transfer. The same applies to the photoionization
rates that occur in the expression for $r$. Thus the set of statistical
equilibrium equations (24) are solved together with the transfer equations
for all these radiative transitions. Such solutions have been discussed
extensively in the literature (e.g., Vernazza, Avrett, \& Loeser 1973,
Mihalas 1978, Avrett \& Loeser 1992) and are not reviewed here. 

Adding the $G_m$ term to each ionization rate $P_{m\kappa}$ in equation (24)
incorporates the effect of flows into the statistical equilibrium equations
for the number densities $n_m$, but $G_m$ has a complex dependence on the
number densities. We have developed the iterative methods of solution
described here to solve for these interdependent quantities.
As in our previous papers we solve the radiative transfer and statistical
equilibrium equations for C, Si, Al, Mg, Ca, Fe, Na and other constituents,
in addition to H and He, to determine the electron density and the various
opacities needed in the radiative transfer calculations. However, our
solutions for these trace species do not include particle flows.


\subsection{Helium Flow}

The ionization balance equations for helium are expressed by two independent
equations of the form of equation (13) for \ion{He}{1} and \ion{He}{2}
particle densities, $n_\alpha$ and $n_\beta$ respectively. A third equation
for fully ionized He (viz., \ion{He}{3} with particle density $n_\gamma$)
would be redundant. As we did in the case of hydrogen, we complement the two
equations for $n_\alpha$ and $n_\beta$ with the equation (9) for the helium total
density, $n_{\rm He} = n_\alpha + n_\beta + n_\gamma$, and use the following
expressions for the net rates of creation of \ion{He}{1} and \ion{He}{2}
\begin{eqnarray}
R_\alpha &=& n_\beta P_{\beta \alpha} - n_\alpha P_{\alpha \beta} \nonumber \\
         &=& (n_{\rm He} - n_\gamma -n_\alpha ) P_{\beta \alpha}
               - n_\alpha P_{\alpha \beta}  \nonumber \\
R_\beta  &=& n_\alpha P_{\alpha \beta} + n_\gamma P_{\gamma \beta}
               -n_\beta ( P_{\beta \alpha} + P_{\beta \gamma} ) \\
         &=& n_\alpha P_{\alpha \beta} + ( n_{\rm He} - n_\beta - n_\alpha )
               P_{\gamma \beta} - n_\beta ( P_{\beta \alpha} + P_{\beta \gamma} )
               \, . \nonumber
\end{eqnarray}
Thus, for helium, two equations (13) apply for $y_\alpha = n_\alpha / n_{\rm He}$
and $y_\beta = n_\beta / n_{\rm He}$, the \ion{He}{1} and \ion{He}{2} fractions,
and with definitions for $r$ and $s$ given by
\begin{eqnarray}
r_\alpha &=& n_{\rm He} ( P_{\alpha \beta} + P_{\beta \alpha} ) \nonumber \\
s_\alpha &=& n_{\rm He} (1 - y_\gamma) P_{\beta \alpha} \nonumber \\
r_\beta  &=& n_{\rm He} ( P_{\beta \gamma} + P_{\gamma \beta}
                         + P_{\beta \alpha} ) \\
s_\beta  &=& n_{\rm He} [ (1 - y_\alpha) P_{\gamma \beta}
                         + y_\alpha P_{\alpha \beta} ] \nonumber
\end{eqnarray}
where $y_\gamma = n_\gamma / n_{\rm He} = 1 - y_\alpha - y_\beta$.

Again assuming $v_{ik} = 0$ we can use the transformation in equation (16) for each
of the two equations with the corresponding definitions of the parameters $r$
and $s$, so that the formal solutions for the case of zero diffusion are
\begin{eqnarray}
y_\alpha (\zeta_\alpha) &=& y_\alpha(0) \exp^{ - {\zeta_\alpha / F_{\rm He} } }
     + \int_0^{\zeta_\alpha} [ \sigma_\alpha (\zeta_\alpha^\prime )
     \exp^{ - {(\zeta_\alpha -\zeta_\alpha^\prime )} / F_{\rm He} } ] 
     { d\zeta_\alpha^\prime / F_{\rm He} } \nonumber  \\
y_\beta (\zeta_\beta)   &=& y_\beta(0) \exp^{ - {\zeta_\beta / F_{\rm He} } }
     + \int_0^{\zeta_\beta} [ \sigma_\beta (\zeta_\beta^\prime )
     \exp^{ - {(\zeta_\beta -\zeta_\beta^\prime )} / F_{\rm He} } ] 
     { d\zeta_\beta^\prime / F_{\rm He} } \, ,
\end{eqnarray}
where
\begin{eqnarray}
\sigma_\alpha &=& ( 1 - y_\gamma ) { P_{\beta \alpha} \over
                    { P_{\alpha \beta} + P_{\beta \alpha} } }  \nonumber \\
\sigma_\beta  &=& { { ( 1 - y_\alpha ) P_{\gamma \beta} + y_\alpha P_{\alpha \beta} } 
                    \over { P_{\beta \gamma} + P_{\gamma \beta} + P_{\beta \alpha} } }
                    \, ,
\end{eqnarray}
and where $F_{\rm He}$ is the total helium particle flow from equation (11).

Note that $\sigma_\alpha$ depends on $y_\gamma$ and that $\sigma_\beta$ depends on
$y_\alpha$. Thus we iterate between the two equations (28) for $y_\alpha$ and
$y_\beta$. We have found that this iteration converges successfully to consistent
values.

From the starting values of $n_\alpha$, $n_\beta$ and $n_\gamma$ as functions of
height we solve equations (28) to obtain new values, maintaining the same
$n_{\rm He}$. The number densities of \ion{He}{1} and \ion{He}{2} levels
are renormalized accordingly.

For \ion{He}{1}, these number densities are used in an equation similar to the
hydrogen equation (25) to determine the coefficients $G_m$ for \ion{He}{1}
level $m$. Then the statistical equilibrium equations, similar to equation (24),
are combined with the radiative transfer equations and solved to get new values
of the number densities $n_m$ for each level of \ion{He}{1}.

The treatment of \ion{He}{2} differs from that of H and \ion{He}{1} since $R_\beta$
in equation (26) includes the rates of ionization from and recombination to a lower
ionization stage. Thus, for the ground level of \ion{He}{2} we use the \ion{He}{1}
ionization equilibrium equation to eliminate \ion{He}{1} ionization and
\ion{He}{2} recombination. For the zero diffusion case we then define
\begin{equation}
G_1 ({\rm HeII}) = { F_{\rm He} \over n_1 } { d \over dz }
                 \left( { n_1 \over n_{\rm He} } + y_\alpha \right)
\end{equation}
for the ground level of \ion{He}{2}, leaving the expression corresponding to
equation (25) unchanged for the \ion{He}{2} levels $m > 1$. Here we assume
that there are no significant direct transitions between \ion{He}{1} and the
excited levels of \ion{He}{2}. With $G_m$ defined in this way, the statistical
equilibrium equation (24) applies in all cases.


\subsection{The Gas Pressure}

In the cases where flows are present we also need to consider how they affect
the momentum balance equation from which we compute the gas pressure and
consequently the H and He particle densities. For this we use the
Navier-Stokes equation, neglecting viscosity and magnetic forces,
\begin{equation}
{ \partial U \over \partial t } + ( U \cdot \nabla ) U = g - { {\nabla p}
             \over \rho } \, ,
\end{equation}
which in the present case reduces to
\begin{equation}
U { dU \over dz } = g - { 1 \over \rho } { dp \over dz } \, ,
\end{equation}
where $g$ is the gravitational acceleration and $p$ is the gas pressure.

Since in the current models $p$ is neither constant nor an analytic
function of $\rho$, this equation has to be integrated numerically to yield the
proper pressure and density as functions of height. We use
the definition of $F_m$ from the mass conservation equation (8) and write
\begin{equation}
F_m { d \over dz } \left( { F_m \over \rho } \right) + { d \over dz }
    \left( \rho { V_{tp}^2 \over 2 } \right) = \rho g - { dp \over dz }
\end{equation}
where we include a Bernoulli term, the ``turbulent pressure'' gradient,
based on the ``turbulent pressure velocity'' $V_{tp}$. As in the FAL papers,
$V_{tp}$ is inferred from the observed non-thermal widths of lines formed
at various heights. Collecting terms in this equation and considering that
$F_m$ is constant gives the following expression that we integrate numerically
\begin{equation}
{ d \over dz } \left( { F_m^2 \over \rho } + \rho { V_{tp}^2 \over 2 } + p \right)
              = \rho g \, .
\end{equation}
The sum of terms in parentheses is the ``total pressure'' $p_{\rm total}$.
The integration of equation (34) is done carefully to assure that the nearly
exponential behavior is properly obtained depending on the mass flow and
temperature variation with height. The first term inside the gradient on the
left hand side is often called ``ram pressure'' and takes into account the
effects of the mass flow velocity on the density stratification of the atmosphere.
Because of the ram pressure term, the static density stratification is not
intermediate between the density stratifications of the inflow and outflow
cases.

A problem with models including mass flow is that when the velocity reaches
the sound speed (and viscosity is neglected) the combination of this equation
with the energy balance equation leads to undefined mathematical conditions
and erroneous numerical results. In this paper we avoid this difficulty by
confining our models to the subsonic regions.


\subsection{Flow Results}

Here we discuss the effects of the flow alone, as obtained from the above
equations that assume $v_{ik} = 0$.

We show the results obtained for different values of $F_{\rm H}$ using a
prescribed $T(z)$ model that has been calculated for a realistic energy
 balance static case (see below). In this section we consider only three
values $F_{\rm H}: +5 \times 10^{15}$ (outward flow), 0 (zero flow), and 
$-5 \times 10^{15}$ particles cm$^{-2}$ s$^{-1}$ (inward flow) and
the corresponding values for $F_{\rm He} = a_{\rm He} F_{\rm H}$, where
$a_{\rm He}$ is the relative helium abundance which in this paper is
assumed to have the constant value $a_{\rm He} = 0.1$. This corresponds
to zero relative velocity between H and He particles. We use the
designations out5$^{\prime\prime}$, 0$^{\prime\prime}$, and
in5$^{\prime\prime}$, respectively, for these models, using the
double prime to indicate that these are models with a prescribed
temperature structure and do not include diffusion.

Figure 1 shows part of the $T(z)$ distribution used in all three sets
of calculations. This is essentially an extended version of the modified
energy-balance static model C (Fontenla et al. 1999) based on FAL3
mentioned earlier.  That model includes particle diffusion. Here we want
to show the effects of flow without diffusion using this prescribed $T(z)$
distribution; hence the calculation in this section (like the equations
shown in the previous section) omits the diffusion terms that are included
in the determination of model C. Thus, the results given in this section
are not realistic but are given for illustrative purposes only. This model
extends up to coronal temperatures of $1.6 \times 10^6$ K, but in Figure 1
we show only the low transition region where H and He are only
partially ionized.

Figure 2a shows $y_{\rm a}(z)$ and Figure 2b shows $y_{\rm a}(T)$ for the
three values of $F_{\rm H}$ considered here.  These results show that outflow
transports low-ionization material outwards while inflow transports
high-ionization material inwards. In the former case, as the material travels
into the high temperature region it becomes ionized by collisions with fast
electrons. In the latter case the ionized material transported into the low
temperature region recombines gradually.

Figures 2c and 2d show that $y_\alpha$ in the three cases has the same general
behavior at lower temperatures as $y_{\rm a}$, but at higher temperatures
$y_\alpha$ is almost 100 times larger for outflow than for inflow.

Figures 2e and 2f show $y_\beta$ in the three cases. In the outflow case, He
is in the form of \ion{He}{2} throughout a very extended region. The results
show an enhancement of \ion{He}{2}, relative to the static case, in the upper
chromosphere for both outflow and inflow; also, the narrow maximum in \ion{He}{2}
at the base of the transition region in the static model disappears in both
moving cases.

The results for \ion{He}{2} are more complex than the others because outflows
transport \ion{He}{2} to higher temperatures but at the same time the ionization
of \ion{He}{1} transported from the chromosphere prevents \ion{He}{2}
depletion. Inward flow drives the \ion{He}{2} and \ion{He}{3} into the
chromosphere where \ion{He}{3} quickly recombines into \ion{He}{2},
while \ion{He}{2} recombines gradually.

We do not discuss these solutions further because they do not include diffusion.
As noted at the beginning of this section, the purpose of these solutions is
only to illustrate how velocity effects alone would affect the ionization
balance.


\section{Combined Flow and Diffusion}

Now we consider the general case including the diffusion velocities,
which were ignored in the preceding sections to simplify the discussion.
The basic equations for the H and He diffusion velocities appear in FAL3,
and here we restate some of them in summary form only.  Again equation (13)
is the basic equation and we use the same expressions as in \S 3 for
$r$ and $s$. The transformation in equation (16) leading to equation (15)
also can be carried out here. However, because $v_{ik}$ is not predefined but
rather depends on the gradient of the ionization fraction, the analytic
solution given by equation (18) is no longer possible. Thus, we replace
$v_{ik}$ by explicit expressions derived from particle transfer theory
(see FAL3, and also Braginskii 1965), and solve equation (13) numerically.

In general, the diffusion velocities can be expressed as linear functions
of the logarithmic gradients of the ``thermodynamic forces'' (FAL1, 2, 3).
This expression is
\begin{equation}
v_{ik} = \sum_n D_{ikn} Z_n \, ,
\end{equation}
where $D_{ikn}$ is the diffusion coefficient that expresses the dependence
of the diffusion velocity of element $k$ in ionization stage $i$ on the
thermodynamic force $Z_n$. In this paper we use the same expressions for $Z_n$
that we gave in FAL3 (see eqns. 15 and 16 in that paper) and
we again neglect gravitational thermodynamic forces. 
As we did in our previous papers we express the relative diffusion velocities
$u_{i,i+1,k}$ between consecutive ionization stages $i$ and $i+1$ of
element $k$ as linear functions of the thermodynamic forces and then express
the diffusion velocities $v_{ik}$ as linear functions of these relative
diffusion velocities.

The definitions of the relative diffusion velocities as well as the
expressions of the diffusion velocities in terms of the relative diffusion
velocities are shown in FAL3 (see eqns. 14 and 17 in that paper). Although
we solve the ionization equations in a slightly different way here,
FAL3 gives the remaining details on the treatment of diffusion.


\subsection{Hydrogen Flow and Diffusion}

The ionization equilibrium equation (13) for H is
\begin{equation}
{ d \over dz } [ y_{\rm a} ( n_{\rm H} v_{\rm a} + F_{\rm H} ) ] 
       + r y_{\rm a} = s \, ,
\end{equation}
and using the derivations in FAL3 (mainly eqns. 16 and 17 in that paper)
we can write
\begin{equation}
v_{\rm a} =  { x \over {1+x} } (d_{11} { {d \ln x} \over dz } + \Delta_1 )
\end{equation}
where $x = n_{\rm p} / n_{\rm a}$, and
\begin{equation}
\Delta_1 = d_{12}Z_a + d_{13}Z_b + d_{14} Z_c + d_{15} Z_T \, .
\end{equation}
It can be shown without difficulty that
\begin{equation}
n_{\rm a} { x \over {1+x} } { {d \ln x} \over dz } 
    = - n_{\rm H} { {d y_{\rm a} } \over dz } \, ,
\end{equation}
where $y_{\rm a} = n_{\rm a} / n_{\rm H}$ as before.
Thus equation (36) becomes
\begin{equation}
{ d \over dz } ( g y_{\rm a} - f { {d y_{\rm a}} \over dz } ) 
      + r y_{\rm a} = s
\end{equation}
where
\begin{eqnarray}
f &=& d_{11} n_{\rm H} \nonumber \\
g &=& F_{\rm H} + n_{\rm p} \Delta_1
\end{eqnarray}
and where $r$ and $s$ are given by equation (20). Appendix A gives the
numerical method we use for solving the differential equation (40).

The solution of these equations is again iterated with the radiative
transfer and statistical equilibrium equations for the level populations
of hydrogen as described above, but now also including the hydrogen atom
diffusion velocity $v_{\rm a}$ in the expression for $G_m$
for hydrogen level $m$ so that
\begin{equation}
G_m = { 1 \over n_m } { d \over dz } [ n_m ( v_{\rm a}
           + { F_{\rm H} \over n_{\rm H} } ) ] \, ,
\end{equation}
which replaces equation (25). Due to the large transition rates between
bound levels, we assume that all the bound levels have the same
diffusion velocity $v_{\rm a}$.


\subsection{Helium Flow and Diffusion}

The ionization equilibrium equations for \ion{He}{1} and \ion{He}{2} are
\begin{eqnarray}
{ d \over dz } [ y_\alpha ( n_{\rm He} v_\alpha + F_{\rm He} ) ]
    + r_\alpha y_\alpha &=& s_\alpha  \nonumber \\
{ d \over dz } [ y_\beta ( n_{\rm He} v_\beta + F_{\rm He} ) ]
    + r_\beta y_\beta &=& s_\beta 
\end{eqnarray}
Using the expressions for the diffusion velocities of neutral and ionized He,
$v_\alpha$ and $v_\beta$, derived in FAL3,
these diffusion velocities are written as linear
functions of the thermodynamic forces and lead to the following equations:
\begin{eqnarray}
{ d \over dz } [ g_\alpha y_\alpha - f_\alpha { d \over dz } y_\alpha ]
                + r_\alpha y_\alpha &=& s_\alpha  \nonumber \\
{ d \over dz } [ g_\beta y_\beta - f_\beta { d \over dz } y_\beta ]
                + r_\beta y_\beta &=& s_\beta 
\end{eqnarray}
where as shown in Appendix B,
\begin{eqnarray}
f_\alpha &=& n_{\rm He} \{ ( 1 + { y_\gamma \over y_\beta } ) 
             [ d_{33} ( 1 - y_\gamma ) - d_{34} y_\alpha ] 
             + { y_\gamma \over y_\beta } [ d_{43} ( 1 - y_\gamma )
             - d_{44} y_\alpha ] \}  \nonumber \\
g_\alpha &=& F_{\rm He} + n_{\rm He} [ (1 - y_\alpha ) \Delta_4
             + y_\gamma \Delta_5 ]  \nonumber \\
f_\beta  &=& n_{\rm He} \{ d_{33} ( 1 - y_\gamma ) - d_{34} y_\alpha
             - { y_\gamma \over y_\alpha } [ d_{43} ( 1 - y_\gamma ) 
             - d_{44} y_\alpha ] \}  \\
g_\beta  &=&  F_{\rm He} + n_{\rm He} [ y_\gamma \Delta_7 
             - y_\alpha \Delta_6 ]  \nonumber
\end{eqnarray}

We again solve these equations using the numerical method described in
Appendix A. Since the coefficients $f$ and $g$ depend on $y_\alpha$, $y_\beta$,
and $y_\gamma$ we iterate between
the solutions of equations (44) for $y_\alpha$ and $y_\beta$ until convergence
is achieved. Also, as in \S 3, we iterate between the solutions of these
equations and the radiative transfer and statistical equilibrium solutions
for the bound levels of \ion{He}{1} and \ion{He}{2}, again introducing
the diffusion velocities $v_\alpha$ and $v_\beta$ in the definitions of the
corresponding $G_m$.


\subsection{Flow and Diffusion Results}

We now show the calculated $y_{\rm a}$ distributions for the same $T(z)$
and the same set of three values of $F_{\rm H}$ (i.e., $+5 \times 10^{15}$
[outward flow], 0 [zero flow], and $-5 \times 10^{15}$ particles
cm$^{-2}$ s$^{-1}$ [inward flow]) considered before but now including
diffusion. We designate these models as out5$^\prime$, 0, and in5$^\prime$,
respectively, using the single prime to indicate that these are models
with the same prescribed temperature structure as before but now include
diffusion. (Models 0$^{\prime\prime}$ and 0 are those without and with
diffusion.) Figures 3a and 3b plot $y_{\rm a}$ vs. $z$ and $T$,
respectively. Comparing Figures 2 and 3 shows that diffusion causes basic
changes in the shape of the calculated curves in the lower transition
region since (as shown in Fig. 4 below) the diffusion velocities are
greater than the flow velocities near $z \sim 2170$ km.

Figures 3c-f show $y_\alpha$ and $y_\beta$ vs. $z$ and $T$. These
\ion{He}{1} and \ion{He}{2} results do not differ greatly from those
in Figure 2 without diffusion. Thus, for this $T(z)$ the effects of
diffusion on the He ionization are not as important as in the case of H.

Our \ion{He}{2} ionization equilibrium calculations include the effects
of dielectronic recombination, based on rate coefficients given by
Romanik (1988). Dielectronic recombination from \ion{He}{2} to
\ion{He}{1} greatly exceeds radiative recombination for temperatures
higher than $10^5$ K, and causes $y_\alpha$ to be more than ten times
larger in this temperature range than would be calculated with
radiative recombination alone.

Figure 4a shows both $U$ and the neutral hydrogen diffusion velocity
$v_{\rm a}$ (eqn. 5) for the three cases. It is clear that in the lower
transition region the \ion{H}{1} diffusion velocity is larger than the
absolute value of $U$ in all cases. Figures 4b and 4c show both $U$ and
the diffusion velocities $v_\alpha$ and $v_\beta$ of \ion{He}{1}
and \ion{He}{2}. For \ion{He}{1} the diffusion velocity is smaller than 
$|U|$, although at some heights not by much. The difference from
the \ion{H}{1} behavior is in part due to the smaller logarithmic
temperature gradient at the heights where the ionization of \ion{He}{1}
to \ion{He}{2} occurs. The \ion{He}{2} diffusion velocities shown in
Figure 4c are much smaller than $|U|$ for a combination of reasons:
because of the large momentum exchange between protons and both \ion{He}{2}
and \ion{He}{3}; and because, at the heights where \ion{He}{2} and
\ion{He}{3} occur, the proton diffusion velocity is small (since H is
fully ionized there).

Figure 5 shows the calculated H Lyman alpha and beta lines at disk center;
Figure 6 shows the \ion{He}{1} resonance lines at 58.4, 53.7, and 1083 nm;
and Figure 7 shows the \ion{He}{2} lines at 30.4, 25.6, and 164 nm.
These figures show that mass flow produces very substantial changes in
these line intensities and profiles far beyond the simple Doppler shifts,
even in the cases considered here, which have same $T(z)$ distribution.
The larger H line intensities in the outflow cases result from the
enhanced neutral H at higher temperatures, producing greater radiative
losses. The converse is true for inflows, although the Lyman $\alpha$
peaks, formed in the chromosphere, are enhanced.

Unit optical depth at the center of the \ion{He}{1} $\lambda$ 58.4 nm line
occurs at the base of the transition region near $z$ = 2160 km, where
$T \approx 10^4$ K.  Figure 3 shows that, in this region, the
values of $y_\alpha$ for both in5$^\prime$ and out5$^\prime$ are smaller
than in the static case and the values of $y_\beta$ for both in5$^\prime$
and out5$^\prime$ are larger than in the static case. The greater helium
ionization in this region leads to enhanced \ion{He}{1} line source
functions.  As Figures 6a and 6b show, this produces emission lines
for both moving models that are stronger than in the static case. The
greater \ion{He}{1} ionization also increases the population of the lower
level of the \ion{He}{1} $\lambda$ 1083 nm line, thus causing greater
absorption of the infrared photospheric continuum.

We return to a consideration of these calculated line profiles in
\S 5 where we show the results for the energy-balance $T(z)$ distribution
in each case.


\section{Energy Balance}

We now address the computation of self-consistent energy balance models of
the transition region. Our methods are similar to those in our earlier papers
but here we add the terms corresponding to the ionization energy and enthalpy
transport due to the particle flows, $F_{\rm H}$ and $F_{\rm He}$, in
addition to the corresponding terms due to the heat conduction and diffusion
velocities already considered in our earlier papers.

As explained in Vernazza et al. (1981, \S {IX}), we use the PANDORA
computer program to compute the radiative losses due to H and He from
the solutions of the equations shown above. These radiative losses now
include the effects of flow as well as those of diffusion included in FAL3.
To these radiative losses we add the free-free and other
elemental radiative losses, using estimates based on the work of Cox and
Tucker (1969). In the upper transition region radiative losses due to elements
other than H and He dominate, and these estimates are only approximate since
they do not take mass flow and particle diffusion into account. However, we
consider our solutions to be accurate in the lower transition region, where
losses due to H and He indeed greatly dominate. We will recompute the upper
transition region of our models in a subsequent paper after obtaining
better estimates of the radiative losses for elements other than H and He
including the effects of particle diffusion and elemental flows.

After computing the total radiative loss, $q_R(z)$, we calculate $F_R(z)$,
the integral of this quantity from the lower boundary of our energy balance
calculation, $z_0$, out to the given height $z$ to obtain the inward energy
flux at $z$ required to compensate for the radiative losses between $z_0$
and $z$.

As in FAL1, 2, and 3, we locate this lower boundary at the top of the
chromosphere where the transition region is assumed to start. The height $z_0$
is obtained from observational constraints and cannot be derived from theory
because it depends entirely on the details of chromospheric heating that are
not yet well understood. Thus, the temperatures at heights below $z_0$ are
those given by our semi-empirical model C described above. We compute the
temperature stratification above $z_0$ by requiring that the downward heat
transport balances the radiative losses (minus any mechanical dissipation,
see below), or equivalently that the decrement of the total heat flux is
equal to the value of $F_R$ mentioned above (minus any mechanical energy).
The values of the height and temperature at our chosen boundary are
$z_0$ = 2163.25 km and $T_0$ = 9530 K. Table 1 lists the full set of
atmospheric parameters for the current version of model C used in the
present paper. (This version differs slightly from the earlier one used by
Fontenla et al. (1999) due the choice of $z_0$, the extension to higher
temperatures, and to various improvements in the PANDORA calculations.)

Our method described here can take into account an ad hoc mechanical
energy dissipation (or heating) term, $q_M$, that is assumed to have the
form
\begin{equation}
q_M = C_q n_{\rm H} \, ,
\end{equation}
where $C_q$ is a coefficient chosen to account for this dissipation. This
mechanical energy dissipation term is also integrated from the lowest
boundary of our energy balance calculation, $z_0$, to obtain the mechanical
energy flux, $F_M$, dissipated between the heights $z_0$ and $z$. 
However, we find that for the cases shown here, introducing $q_M$ as in
equation (46), with the coefficient $C_q$ chosen in such a way as to compensate
for the highly variable upper chromospheric radiative losses, has effects on
lower transition region models that vary in each case. Here we want to show
only the effects of velocity without introducing the further complication of
variable mechanical heating since in many cases (inflows, the static case,
and small outflows) a value of $C_q$ that would account for the upper
chromospheric losses has no significant effect on the lower transition region.
Thus, in this paper we have confined ourselves to cases with $C_q = 0$. We do
not include models with larger outflow in this paper because such cases indeed
require much larger chromospheric heating which would have a
significant effect on the lower transition region. In a later paper we will
consider non-zero values of $C_q$ and show the effects of mechanical heating
on the lower transition region, specially in cases of large outflow.

The total heat flux, $F_h$, is defined here as the sum of heat conduction,
ionization energy transport and enthalpy transport terms as we show below.
The energy balance requirement (from FAL2, eqs. 9-12) takes the form
\begin{equation}
F_E = F_h(z) + F_R(z) - F_M(z)
\end{equation}
where $F_E$ is the constant value that results from the specified lower boundary
condition, and is equivalent to the total heat flux at the lower boundary $z_0$.
This energy balance equation is sometimes formulated by equating to zero the
divergence of the right-hand-side of equation (47) (or the derivative with
respect to $z$ in our one-dimensional modeling), which is equivalent to the
integral form shown here.

The total heat flux $F_h$ can be expressed as
\begin{eqnarray}
F_h &=& F_T + F_U \, ,  \nonumber \\
F_T &=& -\kappa {{ d \ln T} \over dz} + {5 \over 2} n_e k T v_e +
          \sum_{i,k} n_{ik} ({5 \over 2} k T - E_{\rm ion})
          (v_{ik} + v_k) \, , \nonumber \\
F_U &=& {5 \over 2} n_e k T U + \sum_{i,k} n_{ik} ({ 5 \over 2} k T 
          - E_{\rm ion}) U  \, , \\
    &=& { 5 \over 2 } p U - U \sum_{i,k} n_{ik} E_{\rm ion} \, , \nonumber
\end{eqnarray}
where $F_T$ is the thermally-driven heat flux, $F_U$ is the mass-flow-driven
heat flux, $E_{\rm ion}$ is the ionization energy for element $k$ from ionization
stage $i$ to the fully ionized stage, $\kappa$ is the coefficient of heat conduction,
and the other symbols have their previous meaning. (Note that for the heat
conduction we use not the plain gradient of the temperature but the logarithmic
temperature gradient.) In these equations we use the condition of zero electric
current that determines the electron diffusion velocity $v_e$ to be the same as
the sum of the diffusion velocities of all hydrogen and helium ions weighted by
their charge. Since we apply these equations at heights above the chromosphere,
the ions of elements other than H and He can be neglected for the purpose of
computing $v_e$.  The zero electric current condition leads to a
``thermoelectric'' field (see MacNeice, Fontenla, \& Ljepojevic 1991),
which is implicitly included in all our calculations. The terms in $F_h$ that
contain $E_{\rm ion}$ are called ``reactive heat flux'' terms; those in $F_T$
are thermally driven, and those in $F_U$ velocity driven.

The equations for $F_T$ and $F_U$ can be elaborated using the definitions of
$v_e$ and the mass flow $F_m$, so that
\begin{equation}
F_T = F_{\rm cond} + {5 \over 2} k T
          \sum_{i,k} n_{ik} ( v_{ik} + v_k ) q_{i} +
          {5 \over 2} k T \sum_k n_k v_k + F_{T{\rm react}} \, , 
\end{equation}
where $F_{\rm cond} = -\kappa d \ln T / dz$ is the conductive heat flux,
$q_{i}$ is the electric charge of the ionization stage $i$, and where
\begin{equation}
F_{T{\rm react}} = - \sum_{i,k} n_{ik} E_{\rm ion} (v_{ik} + v_k )
\end{equation}
is the thermally-driven reactive heat flux; and
\begin{equation}
F_U = {5 \over 2} {p \over \rho}  F_m  + F_{U{\rm react}} \, , 
\end{equation}
where
\begin{equation}
F_{U{\rm react}} = - U \sum_{i,k} n_{ik} E_{\rm ion}
                 = - (F_m / \rho ) \sum_{i,k} n_{ik} E_{\rm ion} \, , 
\end{equation}
is the velocity-driven reactive heat flux.

Our method for determining the temperature structure using the energy
balance equation (47) assumes that the thermally-driven heat flux $F_T$
can be expressed as a coefficient times the logarithmic temperature gradient.
This is plainly true for the heat conduction term and is approximately
true for the other terms because they depend on the diffusion velocities.
The diffusion velocities are also driven by the temperature gradient:
directly in the case of thermal diffusion, and indirectly in the case of
ionization gradients. Thus in the transition region the dependence of the
thermal heat flux $F_T$ on temperature can be described by
\begin{equation}
F_T = -K(z) {{d \ln T} \over dz}
\end{equation}
where the coefficient $K$ is obtained by dividing $F_T$ (computed in detail
from the equation 49) by the logarithmic temperature gradient. In our
earlier papers we called $K$ the ``effective'' heat transport coefficient.

The first term in equation (51) for $F_U$ varies because the ratio of
gas pressure to mass density, related to the sound speed, varies as the
temperature changes; however, the second term has a stronger variation
because it depends on the H and He ionization changes. Although these
variations are due to the temperature variation, $F_U$ cannot be expressed
as a linear function of the logarithmic temperature gradient. We take
$F_U$ as given in our procedure for correcting the height scale and
recompute it in each iteration of the overall procedure.

After computing the radiative losses, energy fluxes, $K$, and $F_U$,
we adjust the position of each point of our height grid, stretching or
compressing the intervals between adjacent points in such a way that
equation (47) is satisfied at all heights in the transition region.
For this, we start at $z_0$ and step outwards, recomputing the position
of each height point so that the height interval from the adjacent
lower point, $\Delta z$, satisfies the following equation
\begin{equation}
-K_{i-1/2} { {\Delta \ln T} \over \Delta z } + {F_U}_{i-1/2} = F_E
     - ( F_R - F_M )_i - { {\Delta z} \over 2 } ( q_R - q_M ) _{i-1/2}
\end{equation}
where the index $i$ is that of the lower point of the interval
in question, and the values with the index $i-1/2$ are the mean values of
that interval. Note that the sum of the terms on the right hand side is
equal to $F_E - (F_R - F_M)_{i-1/2}$, and that the sum of the terms on
the left hand side is $(F_h)_{i-1/2}$. Thus, this equation is a numerical
approximation of equation (47) evaluated at the center of the interval.
As the calculation proceeds outward to the next interval, the value of
$(F_R - F_M)_{i-1}$ is recomputed incrementally,
\begin{equation}
(F_R - F_M)_{i-1} = (F_R - F_M)_i + \Delta z (q_R - q_M)_{i-1/2} \, . 
\end{equation}

Equation (54) is quadratic in $\Delta z$; in solving it we select the sign
of the square root term in the solution so that $\Delta z$ is positive,
which often implies that a different sign must be chosen depending on the
sign of $F_U$. We avoid the numerical cancellations which can arise when
the velocities are large, or near zero, and use an asymptotic expression
in such situations.

This scheme for computing a revised height grid is nested in a procedure
consisting of the following steps: compute corrections for $\Delta z$;
apply damping to the computed corrections; construct the revised height grid;
and, recompute the fluxes using the same $K$, $F_U$, $q_R$, and $q_M$. This
height grid revision procedure is iterated a few times.

Then, we recompute the ionization and the non-LTE radiative transfer equation
as described in the earlier sections. We recompute the radiative losses,
energy fluxes, and the coefficient $K$, and solve equation (54) again as
described above.  This procedure converges rapidly and has the virtue of
simplicity, since the ``effective heat transport coefficient'' $K$ hides
the complicated dependencies on the temperature gradient.
This method has served us well in building a grid of models incrementally,
enabling us to start the computation of a new model by using another
one with a different particle flux.


\section{Effects of Velocities in the Self-Consistent Models}

We now discuss the results obtained for a set of models illustrating six
cases of hydrogen particle flow, $F_H$: outflows $+2 \times 10^{15}$ and
$+1 \times 10^{15}$ particles cm$^{-2}$ s$^{-1}$; zero flow; and inflows
of $-1 \times 10^{15}$, $-5 \times 10^{15}$, and $-10 \times 10^{15}$
particles cm$^{-2}$ s$^{-1}$. We refer to this sequence of six models by
the names out2, out1, 0, in1, in5, and in10, respectively. Table 2 shows
the logarithmic gradients of $T$ and $n_{\rm p}/n_{\rm a}$ at two
temperature values, $2 \times 10^4$ and $10^5$ K.

Figure 8a shows the calculated $T(z)$ structures in the low transition region,
and Figure 8b shows the $T(z)$ structures at greater heights in the upper
transition region and the low corona. Clearly, the flow velocities considered
here strongly affect the energy-balance temperature structure throughout the
transition region and low corona.

Inflows lead to much smaller temperature gradients due to the much smaller
need (or no need) of thermally-driven
heat transport to support radiative losses. Large
inflow velocities lead to an extremely extended transition region in which
the variation in energy transported down by the mass flow through each
large height interval is dissipated by the radiative losses in that
interval. At such shallow temperature gradients the thermally-driven heat
flux $F_T$ is negligible.  (Fig. 9, below, illustrates this behavior.)

The opposite is true for outflows. The temperature gradient must increase
as the outward mass flow increases so that the inward thermally-driven heat
transport variation can compensate for the large velocity-driven outward
energy flow variation and the radiative losses.

Note that our statements regarding $T(z)$ apply only to the transition
region and lower corona. In coronal layers above those considered here
the effect of mechanical dissipation becomes very important, and the effects
of velocity reverse themselves at heights beyond the temperature maximum in
the corona because then the energy flow by particle outflows
has a direction opposite to that of the temperature gradient.
Thus outflows cause a thinner transition region with the corona closer to the
chromosphere, but then a more extended high corona where the temperature is
over a million degrees. Of course this is true only if the boundary condition
at the top of the chromosphere remains the same and the coronal heating
increases accordingly.

Figure 9 shows the total heat flux, $F_h$, and its velocity-driven component,
$F_U$ (see equation 48). For outflows, $F_U$ is large and positive
but, due to the very steep temperature gradient, it is overpowered by $F_T$ in
the very thin transition region. For small flow velocities (out2 to in1), the
total heat flux is almost the same at temperatures below about $10^5$ K, and
is about $-2 \times 10^5$ ergs cm$^2$ s$^{-1}$ at that temperature, regardless
of $F_U$ (which has the same sign as the particle flow). For large inflows
the curves diverge, and the heat flux at $10^5$ K for the in10 case is about
four times larger (in absolute value) than for small flows.

Figure 10 shows the gas pressure and electron pressure contributions in the
models. In the chromosphere, the gas pressure has a more gradual decrease than
would be expected in hydrostatic equilibrium without the turbulent pressure
term (see equation 34). The gas and electron pressures reach a local minimum
at the top of the chromosphere below the abrupt increase caused by the
temperature increase (that overpowers the density decrease) in the transition
region.  The ``total pressure'' $p_{\rm total}$ in equation (34) decreases
monotonically with height, as shown in Figure 11 for the in1 model.

Figures 12a and 12c show $n_{\rm H}$ and $n_{\rm e}$ as functions
of height for all the models. The differences between these curves
are mainly due to the different $T(z)$ structures of the models.
As shown in Figures 12b and 12d, $n_{\rm H}$ and $n_{\rm e}$ as function
of $T$ are about the same for all the models except near the upper
boundary where the flow approaches the sound speed.

Figures 13a and 13b show the neutral hydrogen fraction, $y_{\rm a}$, as
functions of $z$ and $T$. The variation of $y_{\rm a}(T)$ for $T$ greater
than $\approx 4 \times 10^4$ K in Figure 13b resembles the curves shown
in Figure 3b for models that all use the same $T(z)$ structure as the static
case here. Thus, the changes in $y_{\rm a}(T)$ due to mass flows above the
region of Lyman $\alpha$ formation are not much affected by the temperature
structure. However, at lower temperatures the curves in Figure 13b all tend
to converge, while those in Figure 3b remain well separated, even at $10^4$ K.
This is because the back-radiation in the energy balance cases is less
affected by the flow (as we show below) than in cases where the temperature
structure was prescribed. For the large inflows in Figure 13b diffusion
becomes negligible and $y_{\rm a}$ depends only on temperature.

The relationship between Figure 13a and Figure 3a is more complicated
because Figure 13a combines the effects of both the changing temperature
gradient and the velocities. Figure 8 shows that the temperature
at a given height is smaller for inflows than for outflows, e.g.,
at $z$ = 2170 km, $T$ = 41200 K for in1 and $T$ = 81800 K for out1.
As a result $y_{\rm a}$ at this height is larger for inflows than for
outflows since hydrogen is not as highly ionized. This contrasts with
the results in Figure 3a, based on a common $T(z)$, where $y_{\rm a}$
at a given height is smaller for inflows than for outflows.

Figure 14 shows the total diffusion velocities of H atoms and ions (protons),
$(v_{\rm a}+v_{\rm H}) = V_{\rm a}$ and $(v_{\rm p}+v_{\rm H}) = V_{\rm p}$
(but note that we assume no relative diffusion velocities between H and He,
so that $v_{\rm H} = v_{\rm He} = 0$). These velocities are substantial for
the outflow and the static models; they are still significant for the inflow
model in1, but are negligible for in5 and in10 since the temperature and
ionization gradients are very small in these cases.

Figure 15 shows the helium diffusion velocities $v_\alpha$ and $v_\beta$.
Comparison with Figures 4b and 4c (for $T(z)$ prescribed) shows that the
diffusion velocities are now different because the changes in $T(z)$ tend
to increase He diffusion for outflows, and to decrease it for inflows. 

Figure 16 shows the reactive components of the energy flow (see \S 5)
pertaining to H and He ionization energy transport. We show
$F_{U{\rm react}}$ from equation (52) (dashed lines) due to the mass
flow alone, and the sum
\begin{equation}
F_{\rm react.total} = F_{U{\rm react}} + F_{T{\rm react}}
\end{equation}
representing the ``total reactive energy flow'' which includes both mass
velocity and particle diffusion effects.  Figures 16a and 16b show these
quantities for H as function of $z$ and $T$, respectively. These figures
show that for H the temperature-driven part (due to diffusion) dominates
in the outflow models, static models, and for in1, but is small for in5
and negligible for in10.  The particle diffusion effect is much less
important for \ion{He}{1} as shown in Figures 16c and 16d, and it is
negligible for \ion{He}{2} (Figures 16e and 16f).  The He
temperature-driven reactive energy flux is negligible in most cases
(except for the static model) because it is overbalanced by the
\ion{He}{2} reactive velocity-driven energy flux.

The temperature-driven H reactive energy flux plays a major role in
the low transition region of all the models except in5 and in10.
The values of $F_{U{\rm react}}$ and $F_{\rm react.total}$
are shown in Table 3 for $2 \times 10^4$ K, and in Table 4 for $10^5$ K.
These tables also list the values of $F_h$ and $F_{\rm cond}$ (eqns. 48
and 49) and the height above $z_0$, the base of the transition region,
in each model. In the large inflow cases the enthalpy energy
flow becomes dominant and so large that radiative losses are only able to
dissipate this energy within a layer of large extent. This leads
to very shallow temperature gradients and to very extended transition
regions in these cases.

Figure 17 shows the total radiative losses $q_R$ as functions of $z$ and $T$.
Figure 17a shows that in most cases these radiative losses are sharply peaked in
the transition region, but this peak shifts to greater heights, and broadens, for
inflows. The opposite is true for outflows. Figure 17b shows that the large
inflow cases practically share a common curve (except for some departure at
high temperature) and have a very flat maximum.  In models with small inflow and
with outflow the H and He peaks become bigger and shift to larger temperatures as
the flow increases; this behavior is typical of the effects of particle diffusion.
Also, we note that at temperatures near $10^4$ K the large inflow models, due
to the H contribution, have larger radiative losses than all the others, but
these radiative losses are not very large in absolute value.

Figures 17c and 17d show the radiative losses scaled differently, in ways
commonly used in the literature. They show the same basic behavior as 17b.
At temperatures above $4 \times 10^5$ K the radiative losses shown in
Figure 17c are almost the same for all models, but this is just due to our
assumption that the function $q / {n_{\rm e} n_{\rm H}}$ as defined
by Cox \& Tucker (1969) accounts for the
radiative losses due to all elements other than H and He.
This assumption is probably not accurate because diffusion and flows would
produce the same effects on other species (although if these are ionized the
diffusion effects are probably small). At lower temperatures where the H and
He radiative losses dominate, these functions are different for all the models,
thus showing that none of the customary scaling laws applies for cases
with flows. A different case is that of large inflows for which Figures 17
show that the radiative loss follows a common curve.


\section{Line Profiles}

In this paper we are not concerned with fitting any particular observations
but only with showing the physical effects of the combined diffusion and
mass flow processes on the H and He ionization and line formation. 
We do not show models of active regions or other specific solar features.
These are postponed to a later paper where we include mechanical energy
dissipation which, for simplicity, is not included here. Given these
limitations, we note how our results compare with available
observations of high spectral, temporal, and spatial resolution, to
indicate where our results are generally consistent with the behavior
of observed lines, and to show where some problems still remain.

Figures 18a and 18b show the Lyman-$\alpha$ profiles for the inflow and
outflow cases, respectively. In the inflow cases the lines become slightly
broader and the peak intensities increase with increasing flow, producing
an increase in the integrated line intensity, but asymmetries only start
to become substantial for large inflows (for in10 the blue peak is larger
than the red peak). For inflows the central intensity increases less than
the peaks do and this leads to a larger relative central reversal where
the peak-to-center ratio may reach $\sim 4$. In the outflow cases
the peak intensities increase very little but the central intensity
increases with increasing flow (thus also producing an increase in the
integrated intensity, but for a different reason). For outflows the
relative central reversal becomes smaller and the peak-to-center ratio
may drop as low as 1.2. For outflows the asymmetry remains small and
manifests itself as a slight increase of the red peak as the deepest
point of the central reversal moves very slightly to the blue. 
This asymmetry is so small that it may be hard to detect in observations.

These changes in line profiles arise for different reasons in inflows
and in outflows. For inflows the diffusion effects become smaller as the
flow increases, and for large inflows become insignificant. This drives the
region of formation of the line center deeper in the atmosphere where
the temperature is lower than in the static case but where the line
source function is enhanced because the electron density is higher,
thus producing a moderate decrease at line center, except for model in10
where there is a moderate increase at line center. The peaks form at
the top of the chromosphere (at $\sim 10^4$ K or less) where diffusion
effects are small in all cases; thus, the increased peak intensities
are just due to the larger electron density and amount of material
at this temperature. For outflows, on the other hand, diffusion effects
increase and shift the region of formation of line center to a layer
which is thinner but has temperatures of $\sim 3 \times 10^4$ K or
more, while the line peaks still form at about the same temperature
and electron density as in the static case.

This behavior of the computed Lyman-$\alpha$ line is consistent with the
very detailed observations by Fontenla, Reichmann, \& Tandberg-Hansen
(1988, hereafter FRT), who found that the relative depth of the central
reversal of this line changes with position (unfortunately, appropriate
time sequences were not available). While the average quiet Sun profile
and absolute intensity are in reasonable agreement with the static models
(see FAL1), FRT showed that some quiet Sun profiles have deeper and
others have shallower central reversals than the average. This effect
is larger in active regions where intense peaks are sometimes observed
at some locations, while at other locations in active regions 
the central reversal of the line is almost filled in. Of course in active
regions the line is generally more intense (see FRT and FAL3) since their
chromospheric temperature, and heating, are larger. Regardless of these
intensity differences, the processes described here, whereby the 
the peak-to-center ratio grows larger with inflows and decreases with
outflows, appear to be consistent with observations. (For example,
Fig. 6 of FRT shows two profiles whose central reversals are very
different, and in Fig. 8 of FRT the inflow profile c has a much deeper
reversal than the outflow profiles b and d.)

Figures 18c and 18d show the Lyman-$\beta$ profiles for the inflow and
outflow cases, respectively. The outflow profiles (Fig. 18d) show increasing
intensity as the flow increases. This is again due the increased effects of
diffusion that drive the line center formation region up to temperatures
of $\sim 3 \times 10^4$ K; the effect of this higher temperature
overpowers the reduction in the thickness of the region of line formation.
The inflow profiles (Fig. 18c), instead, show a slight decrease of intensity
for small inflows due to the shift of the line center formation region
toward lower temperatures because of the reduction of diffusion effects.
With large inflows the line becomes stronger and wider than in the static
case due to the increasing optical thickness of the line-emitting region.
There are no pronounced asymmetries but clearly the line center is shifted
slightly to the red with inflows and to the blue in the static case,
while the outflow cases show increasing blue shift.

The Lyman-$\beta$ line profiles observed by SOHO (Warren, Mariska \& 
Wilhelm 1998) show a peak intensity of $\sim 2000$ erg cm$^{-2}$
s$^{-1}$ sr$^{-1}$ \AA$^{-1}$ and integrated intensity of $\sim 900$
erg cm$^{-2}$ s$^{-1}$ sr$^{-1}$ (or $\sim 800$ when the local continuum
is removed), and these values are consistent with our current static model.
However, the Lyman-$\beta$ profile observed by SOHO shows a very small
but definite asymmetric self-reversal that is not predicted by our present
static calculations. This small central reversal contrasts with larger
ones reported previously from OSO 8 data; however, the large reversals
observed by OSO 8 may have been caused by geocoronal absorption, given
the low orbit of that spacecraft. The relatively low spectral resolution
of these data makes it difficult to separate the geocoronal absorption in
the way used by FRT for Lyman-$\alpha$ line profiles. In the context of our
models, several possibilities remain to account for the small Lyman-$\beta$
central reversal observed by SOHO.  1) One likely possibility suggested
by FRT and by Fontenla, Fillipowski, Tandberg-Hansen, \& Reichmann (1989)
is absorption by a ``cloud layer'' consisting of \ion{H}{1} dynamic
material in the lower corona just above the transition region. The
spicules observed in H$\alpha$ are the densest component of such a layer
but, of course, much more material is visible at the limb in Lyman-$\alpha$
than in H$\alpha$, and all this material would absorb in Lyman-$\beta$.
2) Another likely possibility is that portions of the
disk are covered by regions of inflow, of outflow, and of stationary
material, and that a combination of our computed profiles in these cases
may produce an apparent self-reversal. 3) It is possible that an unknown
blend may cause an apparent self-reversal. However, all higher
Lyman lines show a similar, but decreasing, asymmetric self-reversal
and this is hard to explain by a blend unless it involves a series
analogous to the Lyman series. Consequently we consider a blend
near line center of Lyman-$\beta$ only a remote possibility.
Note that any of the above possible explanations also may account
for the asymmetry of the central reversal and its evolution among
the higher members of the Lyman series.

The \ion{He}{1} line profiles are shown in Figure 19. The behavior of the
resonance line with the lowest excited level energy, the 58.4 nm line, is
somewhat similar to that of Lyman $\alpha$. With increasing inflow the 
self-reversal becomes more pronounced because the peaks increase more than
the line center. However, in contrast with Lyman $\alpha$,
with increasing outflow the line center increases very strongly and
changes the line's shape from self-reversed to almost pure emission
(for out2). The 53.7 nm line behaves similarly; however, 1) it does
not have a self-reversal in the static case but a flat top instead,
and 2) it develops only a moderate self-reversal with increasing inflow.
The infrared absorption line at 108.3 nm deepens with the magnitude of the
flow velocity. This strengthened absorption is mainly due to increased
optical thickness: for inflows, this is caused by the greater extent
of the lower transition region; for outflows, this occurs at the top of
the chromosphere where increased UV radiation enhances the population
of the triplet ground level due to greater \ion{He}{2} recombination
(see FAL3 and Avrett, Fontenla, \& Loeser (1994) for a discussion of 
how this line is formed).

The \ion{He}{2} lines shown in Figure 20 behave differently from the H and
\ion{He}{1} lines. However, their intensities still increase with increasing
flow velocity, regardless of its sign, except in the case of model in1.
The lowest excitation resonance line, at 30.4 nm, has a flat top in the static
case and develops a small self-reversal as the magnitude of the flow velocity
increases. For inflows this occurs because, with increasing velocity, the
line-center-forming region grows thicker geometrically and moves to lower
temperature and higher electron density. For outflows the line intensity
increases strongly because the line-forming region occurs at higher
temperatures and this more than compensates for the smaller optical depth.

The \ion{He}{2} 25.6 nm line also grows stronger with both increasing
inflows and increasing outflows, and becomes rather bright for outflows.
This line has a typical pure emission profile in all cases. Its center shifts
appreciably (even more than Ly $\beta$) since this line forms at greater
heights where velocities are larger.

The \ion{He}{2} 164.0 nm line grows stronger with both increasing inflows
and increasing outflows, but the line shape is flatter for outflows. There
are differences between our computed profiles and observations (e.g.,
Kohl 1977): in our calculations the composite red peak is somewhat larger
than the composite blue peak, while the opposite is generally observed.
Walhstrom \& Carlsson (1994) calculated the 164.0 nm line using model C
of Vernazza et al. (1981) without diffusion or velocities. They found
that the collisional coupling between the three n = 2, as well as the
five n = 3 fine-structure levels, at transition region densities, is
not large enough to populate each group according to their statistical
weights, and that this results in changes in the shape of the composite
164.0 nm line that lead to closer agreement with the observations.
In the present paper, however, we have assumed for simplicity that the
two groups of sublevels are populated according to their statistical
weights, and this may be the reason for the discrepancy with the
observations. Also, since this composite line is broad, comparison
with observations must consider blends with other lines, e.g. 
\ion{Cr}{2} 164.0364 nm, \ion{Fe}{4} 164.037 nm, \ion{Fe}{3} 164.0384 nm,
and possibly other identified lines, that may blend into the
composite red peak.

Our calculations show that, for the moderate flow velocities we consider
here, line shifts are so small that they might only be detectable in
the relatively weak Lyman-$\beta$ line and the somewhat stronger
\ion{He}{2} 25.6 nm line. For the largest inflows, our computed blue peaks
of the bright and saturated or self-reversed resonance lines of H and He
are brighter than the red peaks.  However, for small flows both computed
peaks are nearly equal and the differences would be undetectable.
Consequently the line shift and asymmetry of these bright lines would
not be practical diagnostics of small flows. Rather, the magnitude of
their self-reversal, and their relative intensities, would be better
diagnostic tools.


\section{Discussion}

The present paper differs from the work of CYP in that they do not include
the effects of particle diffusion and velocities in the calculations of H
and He ionization. Consequently their calculated level populations and
radiative losses also are not consistent. Instead, they use values of the
ionization and radiative losses from a previous paper by Kuin and Poland
(1991) that does not include particle diffusion or flow velocities. As a
result, although the CYP paper is an improvement over previous work that
did not consider partial ionization or optically thick radiative losses,
it does not present results from consistent calculations such as the ones
we show here. Despite these shortcomings, CYP provides the correct
qualitative behavior (like earlier papers) in that downflows smooth the
temperature gradient while upflows sharpen it. Quantitatively, however,
most of their results are very different from ours.

The radiative losses presented by Kuin and Poland and used by CYP consider
radiative transfer effects in a static case but greatly underestimate
the H and He radiative losses because they ignore particle diffusion. As
explained in the FAL papers, particle diffusion produces larger H and He
losses than those shown by CYP in their Fig. 1, and these losses occur at
higher temperatures than where these elements emit their line radiation in
models ignoring diffusion. Also, as explained in the FAL papers, our
computed absolute Lyman-$\alpha$ profile is consistent with observations
while calculations like those of Kuin and Poland do not give radiative
losses large enough to correspond to the observed emission in the core of
the Lyman-$\alpha$ line because they do not include particle diffusion.

Furthermore, the consistent calculations shown here demonstrate that the
H and He radiative losses are influenced by the flows.
This is easily seen when comparing Fig. 1 of CYP with Fig. 17
of this paper. In our present calculations the radiative losses for
various flows (at comparable temperatures) differ substantially from
each other except for large downflows. We have shown that this greatly
affects the energy-balance $T(z)$ stratification and the values of the
various components of the heat flux.

Similar considerations apply to the H and He ionization. As explained in the
FAL papers, particle diffusion has an important effect on the ionization
which is not considered by Kuin and Poland and consequently not by CYP.
This is closely related to the statements above about the radiative losses
and we refer the reader to the FAL papers for details. Also, particle flows
have large effects on the ionization and these are ignored by CYP. Because
of the different ionization, the reactive heat transport is different in
our models than in those of CYP.

The net result of all this is that in many cases our $T(z)$ distributions
are much shallower that those of CYP, and that our values of the 
contributions to the heat flux differ from theirs. Since these two
calculations use different boundary conditions and different underlying
chromospheres and photosphere, we do not give detailed numerical comparisons;
but comparisons of their plots and ours show evident differences.

The boundary parameters at the base of our calculation of the transition
region are: density, temperature, H and He ionization, He abundance,
temperature gradient, H and He ionization gradients, H and He particle
flows, and turbulent pressure. Magnetic topology can be added
when a magnetic field structure more complicated than a vertical magnetic
field is considered. The boundary parameters we use are all at the top of
the chromosphere. In principle we could instead choose to specify these
parameters at the coronal boundary.  While boundary conditions can be
prescribed in many ways, it is often easier to compute models by fixing
them at just one of the boundaries. The results obtained in this way,
then, define the values of the parameters at the other boundary.
Thus one can adjust the prescribed chromospheric boundary conditions
to match coronal observations as well.

We compute the energy balance only in the transition region, where we
assume that radiative losses are balanced mainly by the total heat flux
from the corona (including the ionization energy flow, also called
reactive heat flow) and by the enthalpy flow (outward or inward, depending
on the particle flow). Note that the radiative losses (expressed per unit
volume, unit mass, or any other usual form---see Fig. 17) in the transition
region are orders of magnitude larger than those in the underlying
chromosphere or the overlying corona. Thus if one assumes that the
mechanical energy dissipation in the transition region is comparable to
that above and below, one concludes that mechanical energy dissipation
in most cases would have only minor effects in the transition region.
In our calculation of the transition region we therefore set mechanical
energy disspation equal to zero (but this is just a simplifying
assumption since some dissipation is likely).

However, it is essential to include mechanical dissipation when computing
the temperature structures of the chromosphere and the corona (regions which
indirectly affect the transition region as well).  If energy dissipation
were included in our calculations (even in the primitive form of eqn. 46),
then the onset of the large temperature gradient would move
to a greater height, and this gradient itself would be reduced.
Such energy dissipation would partially compensate for the enthalpy
flow in the case of strong outflow. Thus, if we included energy
dissipation, we should be able to obtain solutions for larger
outflows than those considered in this paper. We plan to address
this in subsequent papers. 

Also, we do not extend our current models high into the corona since
without mechanical dissipation the radiative and conductive losses
require the temperature to rise unchecked. Mechanical energy
dissipation balances the radiative and conductive
losses at coronal temperatures, allowing the calculation of loop models
that reach a maximum temperature. Also, if sufficient heating is provided
to balance the radiative losses in the upper transition region, it
would be possible to compute models of cool loops like those studied by
Oluseyi et al. (1999). The Oluseyi et al. paper and the others cited
therein provide a good introduction to the extensive literature on
energy balance in coronal loops; the present paper applies to regimes
that occur only at the footpoints of such loops.

As an example of an observation showing a system of ``cool'' coronal loops
we mention the famous group of large loops observed at the limb in \ion{C}{4}
by the UVSP instrument on board the SMM spacecraft. These loops were the
logo of several publications and were studied by
Fontenla, Filipowski, Tandberg-Hanssen, \& Reichmann (1989)
as an example of large dynamic loops at the solar
limb. As that paper discusses, and as was later confirmed by X-ray
observations from the SMM, the loop system appears to be formed by
condensation of material ejected in a previous small C flare at one of
the footpoints. However, by the time the observation was made, the heating
processes had subsided and the loops were cooling down, likely due mostly
to radiation because of the large density and relatively low temperature
(estimated at $\approx 10^5$ K). However, as the loops cooled down the
material started falling (at a velocity estimated at
$\approx 10$ km s$^{-1}$) and this flow carried enthalpy (and ionization
energy) down with it. Such energy downflow contributed to the cooling at
the top, and to the heating of the legs and feet of the loops. Persistent
low level brightening was observed at the visible foot (the other was
hidden by the disk) and this may be explained by the downward energy
transport. Since this loop was seen for many minutes with little change,
it is likely that quasi-steady conditions prevailed at the legs and
footpoints while the density slowly decreased at the top. Our models
would apply to the footpoints of loops like these, where the velocities
vary on time scales of many minutes. Small scale phenomena of this type
are common on the solar surface even in the quiet Sun and they may be
related to the H Lyman-$\alpha$ spicules and macrospicules (since these
may just be the extended legs of dynamic coronal loops).

An observational test of the results presented here would be to compare
high spectral resolution profiles of various lines obtained simultaneously
for the same high-resolution spatial element at a footpoint of a coronal
loop and to verify how the features of the various lines relate to each
other and to the results shown here. We note that a particular observation
may not match any of the results shown here because it may correspond to
different boundary conditions than those we chose, corresponding to our
model C. However, the trend of changes in line profiles and intensities of
a set of lines of H and He, which may correspond to different flow
velocities, could be compared with the present calculations.

Determining $T(z)$ based on energy balance is simpler in the transition
region than in the underlying upper chromosphere.  Energy balance models
of the upper chromosphere must consider MHD effects that are probably
negligible in the transition region. Also there are other complicating
factors that make the chromospheric problem very difficult. One of
these is the need for accurate estimates of the radiative losses in the
optically thick regime; such loss estimates need to account for the
effects of velocities, and, in many cases, for the effects of
time-dependent flows.

Another complication for the calculation of the chromospheric
structure is that mechanical energy dissipation is expected to depend
on height, temperature, density, velocity, ionization, and magnetic field.
However, until the mechanism of this mechanical dissipation is identified,
these dependencies are also unknown (and there are too many possibilities
to explore). By using a parametric formula like equation (46)
one can only obtain a rough ad hoc estimate of $C_q$ that corresponds
approximately to the radiative losses of model C at some heights in the
chromosphere. However, using equation (46) to compute energy balance
models of the chromosphere may be practically meaningless since a single
constant cannot account for the strong dependencies on atmospheric
parameters which are characteristic of any likely physical mechanism
of chromospheric heating.

A further complication for the determination of the chromospheric structure
is the likelihood of elemental abundance variations, caused by gravitational
settling or electric fields. Solar wind measurements provide indications of
such variations: in solar and stellar winds the abundances of elements with
high first ionization potential (FIP) including He, differ from photospheric
values (Meyer 1996).  Just where in the solar atmosphere these abundance
variations occur is currently not known but the chromosphere is a good
candidate. The study by Hansteen, Leer, \& Holzer (1997), discussed
earlier, included a consideration of the variation of the abundance of
helium relative to hydrogen, starting in the chromosphere.

We believe that the problem of chromospheric heating must be addressed by
proposing plausible physical mechanisms. For a given mechanism we would need 
to: 1) determine the dependence of the heating on the physical parameters,
2) compute the chromospheric structure resulting from the balance between
this heating and the radiative losses, and 3) compare the predicted spectral
signatures with observations from the entire range of heights of the
chromosphere.

Although such a self-consistent approach is not simple, we believe that
current computing resources are sufficient to attempt it.  As we have shown
in our papers on particle diffusion, and here by our modeling of velocities
in the transition region, consistent modeling of physical processes is a
very powerful tool. It can produce results which compare well with
observations and which explain features that are difficult to understand
from oversimplified arguments. Often the explanations are simple once the
main process is understood, as, for example, in the case of the formation
of the H Lyman $\alpha$ profile.


\section{Concluding Remarks}

We have presented here a fully self-consistent treatment of the
radiative transfer, statistical equilibrium, and energy and momentum
balance for the solar transition region that includes steady-state mass
flows as well as particle diffusion. The detailed calculations are
carried out for H and He, while other elements are not treated in a
fully self-consistent manner. However, these other elements have only
a minor influence in the lower transition region, at temperatures
between $10^4$ and $10^5$ K. We will address the effect of these
elements on the upper transition region in a subsequent paper.

We have shown results for various inward (or downward) and outward
(or upward) particle flows. The cases shown here are by no means
exhaustive, and because of the large number of possible boundary
conditions it is beyond the scope of this paper to include
a grid of models. Instead, we have just presented several cases with
the emphasis on showing how the various processes affect the results
and on how the flow velocities affect the emitted spectral lines of
H and He.

Our calculated line intensities and profiles are generally consistent
with the available observations (except for some details as discussed
above). However, detailed comparison with observations would require
custom adjustment of the boundary conditions for various separate spatial
components, and perhaps even treating multi-component models. Given
the limited spatial resolution of many available observations, we need
to combine the calculated spectra from various components for
comparison with observations.  All this is beyond the scope of the
present paper.

We now comment on some of the generic results from our calculations. The
progressive intensity increase of the Lyman-$\alpha$ peaks and increase
of the relative depth of the central reversal (due to the much smaller
line center increase) with increasing inflow can be used as a diagnostic
for small inflows that do not produce appreciable line asymmetry.
Also, the filling up of the line center with little changes to the
line peaks, and consequent reduction of the relative line reversal,
can be used as diagnostics of small outflows. Flows also affect the
Lyman $\beta$ and especially the \ion{He}{2} 25.6 nm peak positions
in a way that could be detected; these line intensities change less
due to flows than due to different solar features. (Compare the
present profiles with those in FAL3 showing various solar features.)

The intensity ratios between H and He lines may well vary due to He
abundance variations, and departures from the results shown here can
be used to estimate these abundance variations. However, as we have
shown, small flow velocities can affect these ratios. Thus, an
analysis of these line ratios must include more than just a few lines
since both the flow effects and the He abundance variation effects
must be disentangled.

We believe the key to understanding the physical mechanisms in the
solar chromosphere, transition region and low corona, and how they relate
to the formation of solar spectra, is not in assessing one particular
observation, or one mean profile, but rather in observing the complete
range of such features that are present in the Sun. In this way one can
evaluate not only the mean spectra but also the variability of the spectra
vs. position and time for various solar features. Such studies together
with the theoretical modeling of physical processes, and especially
MHD processes (which have not yet been simulated in coupling the
upper chromosphere with the low corona), would provide essential
understanding of chromospheric and coronal heating as well
as the origin of the solar wind.

We hope that the computations and methods shown here are a step in
that direction and will encourage more self-consistent calculations
and comparisons with observations. Such studies are badly needed for the
chromosphere (especially the upper chromosphere), where it is essential
to improve upon the rough, unphysical approximations used so far,
and instead to treat MHD processes with realistic computations of the
ionization, excitation, and radiative losses.


\acknowledgments

This research has been supported in part by NASA Grant NAG5-9851.


\appendix


\section{Five-Diagonal Solution of the Second-Order Equation}

The equation to be solved for $y(z)$ is
\begin{equation}
{ d \over {dz} } ( gy - f { {dy} \over {dz} } ) + ry = s \, ,
\end{equation}
where $f$, $g$, $r$, and $s$ are assumed to be given at the discrete values
of $z_i, i = 1, 2, \cdots, N$. We write the derivatives of any function
$y(z)$ at depth $i$ as
\begin{equation}
y_i^\prime = { { y_{i+1} - y_{i-1} } \over { z_{i+1} - z_{i-1} } } \, ,
\end{equation}
and in this manner obtain
\begin{eqnarray}
(gy - fy^\prime)_i^\prime &=& { 1 \over d_i } [ (gy - fy^\prime)_{i+1}
                                - (gy - fy^\prime)_{i-1} ] \nonumber \\
  &=& { 1 \over d_i } \biggl\{ [ g_{i+1}y_{i+1} - f_{i+1}(y_{i+2} - y_i)/d_{i+1}] \\
  &   & \phantom{ 1 }
        - [ g_{i-1}y_{i-1} - f_{i-1}(y_i - y_{i-2}) / d_{i-1} ] \biggr\} \nonumber
\end{eqnarray}
where $d_j = z_{j+1} - z_{j-1}$.

Equation (A1) then becomes
\begin{equation}
E_i y_{i-2} + D_i y_{i-1} + C_i y_i + B_i y_{i+1} + A_i y_{i+2} = s_i
\end{equation}
where,
\begin{eqnarray}
E_i &=& - (f_{i-1}/d_{i-1}) / d_i \, , \nonumber \\
D_i &=& - g_{i-1} / d_i \, , \nonumber \\
C_i &=& r_i + ( f_{i-1} / d_{i-1} + f_{i+1} / d_{i+1} ) / d_i \, , \\
B_i &=& g_{i+1} / d_i \, , \nonumber \\
A_i &=& - (f_{i+1}/d_{i+1}) / d_i \, , \nonumber 
\end{eqnarray}
for $i = 3, 4, \cdots , N-1$.

For $i = 1$ and $2$ we let
\begin{equation}
y_1^\prime = (y_2 - y_1) / \Delta_1
\end{equation}
where $\Delta_1 = z_2 - z_1$, and for $i = N-1$ and $N$ we let
\begin{equation}
y_N^\prime = (y_N - y_{N-1} ) / \Delta_N 
\end{equation}
where $\Delta_N = z_N - Z_{N-1}$.  As a result,
\begin{eqnarray}
E_1 &=& D_1 = 0 \nonumber \\
C_1 &=& r_1 - ( g_1 + f_1/\Delta_1 - f_2/d_2 ) / \Delta_1 \nonumber \\
B_1 &=& ( g_2 + f_1 / \Delta_1 ) / \Delta_1 \nonumber \\
A_1 &=& - ( f_2/d_2 ) / \Delta_1 \nonumber \\
E_2 &=& 0 \nonumber \\
D_2 &=& - ( g_1 + f_1 / \Delta_1 ) / d_2 \nonumber \\
C_2 &=& r_2 + ( f_1 / \Delta_1 + f_3 / d_3 ) / d_2 \nonumber \\
B_2 &=& g_3 / d_2 \nonumber \\
A_2 &=& - (f_3/d_3) / d_2 \nonumber \\
E_{N-1} &=& - (f_{N-2}/d_{N-2}) / d_{N-1} \\
D_{N-1} &=& - g_{N-2} / d_{N-1} \nonumber \\
C_{N-1} &=& r_{N-1} + ( f_{N-2} / d_{N-2} + F_N / \Delta_N ) / d_{N-1} 
            \nonumber \\
B_{N-1} &=& ( g_N - f_N / \Delta_N ) / d_{N-1} \nonumber \\
A_{N-1} &=& 0 \nonumber \\
E_N &=& - (f_{N-1}/d_{N-1}) / \Delta_N \nonumber \\
D_N &=& - ( g_{N-1} - f_N / \Delta_N ) / \Delta_N \nonumber \\
C_N &=& r_N + ( g_N + f_{N-1} / d_{N-1} - f_N / \Delta_N ) / \Delta_N 
            \nonumber \\
B_N &=& A_N = 0 \nonumber
\end{eqnarray}
The coefficients in equation (A4) then can be determined from the values of
$g$, $f$, $r$, $s$, and $z$, and this five-diagonal set of equations can be solved
for $y_i, \, i = 1, 2, \cdots , N$.

Note that our use of equation (A6) implies a choice of boundary conditions and
leads to a complete specification of the coefficients. This is appropriate
when the divergence of the particle flow and diffusion velocities are negligible.
When there is substantial inflow or outflow we assume that the first term in
equation (A1) is zero at the upstream boundary, so that $y = s/r$ at this boundary.
Thus for inflow we let $C_1 = r_1$, $B_1 = 0$, and $A_1 = 0$, and for outflow
we let $C_N = r_N$, $D_N = 0$, and $E_N = 0$.


\section{Helium Mass Flow with Diffusion}


\subsection{\ion{He}{1}}

Equation (44) for \ion{He}{1} is
\begin{equation}
{ d \over dz } (g_\alpha y_\alpha - f_\alpha { d y_\alpha \over dz })
 + r_\alpha y_\alpha = s_\alpha \, .
\end{equation}
From eqn. (17) of FAL3,
\begin{equation}
v_\alpha = v_{\rm He} + (y_\beta + y_\gamma) V_C + y_\gamma V_D
\end{equation}
where
\begin{eqnarray}
v_{\rm He} &=& y_\alpha v_\alpha + y_\beta v_\beta + y_\gamma v_\gamma \, ,
               \nonumber \\
V_C &=& \Delta_2 + d_{33} { d \over dz } \ln { y_\beta \over y_\alpha }
          + d_{34} { d \over dz } \ln { y_\gamma \over y_\beta } \, ,
               \nonumber \\
V_D &=& \Delta_3 + d_{43} { d \over dz } \ln { y_\beta \over y_\alpha }
          + d_{44} { d \over dz } \ln { y_\gamma \over y_\beta } \, , \\
\Delta_2 &=& d_{31} Z_x + d_{32} Z_a + d_{35} Z_T \, , \nonumber \\
\Delta_3 &=& d_{41} Z_x + d_{42} Z_a + d_{45} Z_T \, . \nonumber
\end{eqnarray}
Expanding each $\ln (x/y)$ as $(\ln x - \ln y)$ and replacing $d\beta / dz$
by $(-d\alpha / dz - d\gamma / dz)$ leads to the result
\begin{eqnarray}
V_C &=& \Delta_4 - ( { {d_{33} - d_{34} } \over y_\beta }
         + { d_{43} \over y_\alpha } ) { d y_\alpha \over dz } \, , \nonumber \\
V_D &=& \Delta_5 - ( { {d_{43} - d_{44} } \over y_\beta } 
         + { d_{43} \over y_\alpha } ) { d y_\alpha \over dz } \, , \\
\Delta_4 &=& \Delta_2 - [ d_{33} { y_\gamma \over y_\beta } 
         - d_{34} ( 1 + { y_\gamma \over y_\beta } ) ] 
         { d \over dz } \ln y_\gamma \nonumber \\
\Delta_5 &=& \Delta_3 - [ d_{43} { y_\gamma \over y_\beta } 
         - d_{44} ( 1 + { y_\gamma \over y_\beta } ) ] 
         { d \over dz } \ln y_\gamma \, . \nonumber 
\end{eqnarray}
Thus in equation (B1) we have
\begin{equation}
g_\alpha = F_{\rm He} + n_{\rm He}
  [ (y_\beta + y_\gamma) \Delta_4 + y_\gamma \Delta_5 ] \, ,
\end{equation}
and
\begin{eqnarray}
f_\alpha &=& n_{\rm He} \biggl\{ (1 + { y_\gamma \over y_\beta })
  [ d_{33} (1 - y_\gamma) - d_{34} y_\alpha ]  \\
  &+& { y_\gamma \over y_\beta } [ d_{43} (1 - y_\gamma) - d_{44} y_\alpha ]
  \biggr\} \, . \nonumber
\end{eqnarray}

We then solve equation (B1) for $y_\alpha$ by the method described in
Appendix A.


\subsection{\ion{He}{2}}

Equation (44) for \ion{He}{2} is
\begin{equation}
{ d \over dz } (g_\beta y_\beta - f_\beta { d y_\beta \over dz }) 
  + r_\beta y_\beta = s_\beta \, .
\end{equation}

From eqn. (17) of FAL3,
\begin{equation}
v_\beta = v_{\rm He} - y_\alpha V_C + y_\gamma V_D \, .
\end{equation}
Expanding the logarithmic terms in equation (B3) as before, but now replacing
$d y_\alpha / dz$ by $(- d y_\beta / dz - d y_\gamma / dz)$ leads to the result
\begin{eqnarray}
V_C &=& \Delta_6 + { 1 \over { y_\alpha y_\beta} } [ d_{33} (1 - y_\gamma)
        - d_{34} y_\alpha ] { d y_\beta \over dz } \, , \nonumber \\
V_D &=& \Delta_7 + { 1 \over { y_\alpha y_\beta} } [ d_{43} (1 - y_\gamma)
        - d_{44} y_\alpha ] { d y_\beta \over dz } \, , \nonumber \\
\Delta_6 &=& \Delta_2 + ( d_{33} { y_\gamma \over y_\alpha } + d_{34})
        { d \over dz } \ln y_\gamma \, , \\
\Delta_7 &=& \Delta_3 + ( d_{43} { y_\gamma \over y_\alpha } + d_{44})
        { d \over dz } \ln y_\gamma \, . \nonumber 
\end{eqnarray}
Finally in equation (B7) we have
\begin{equation}
g_\beta = F_{\rm He} + n_{\rm He} ( y_\gamma \Delta_7 - y_{\rm a} \Delta_6) \, ,
\end{equation}
and
\begin{equation}
f_\beta = n_{\rm He} \left\{ d_{33} (1 - y_\gamma) - d_{34} y_\alpha 
   - { y_\gamma \over y_\alpha } [ d_{43} (1 - y_\gamma) - d_{44} y_\alpha ] \right\}
\end{equation}
and we solve equation (B7) for $y_\beta$ in the same way as described in Appendix A.
Note that $g_\beta$ and $f_\beta$, as well as $r_\beta$ and $s_\beta$ depend on the
values of $y_\alpha$ obtained from the \ion{He}{1} solution, which, in turn depends
on the \ion{He}{2} solution, but to a lesser extent. Thus we solve equations (44)
for $y_\alpha$ and $y_\beta$ iteratively, until consistent values are obtained.


\clearpage

\begin{deluxetable}{cccccccc}
\tablecaption{ {\sc The Current Version of Model C Used in the Present
              Calculation } \label{TAB1} }
\tablewidth{0pt}
\tablecolumns{8}
\tablehead{
\colhead{Depth} &
\colhead{ } &
\colhead{ } &
\colhead{Turbulent} &
\colhead{Hydrogen} &
\colhead{Electron} &
\colhead{Gas} &
\colhead{Total} \\
\colhead{index} &
\colhead{Height} &
\colhead{Temperature} &
\colhead{velocity} &
\colhead{density} &
\colhead{density} &
\colhead{pressure} &
\colhead{pressure} \\
\colhead{$i$} &
\colhead{$z$} &
\colhead{$T$} &
\colhead{$V_{tp}$} &
\colhead{$n_{\rm H}$} &
\colhead{$n_e$} &
\colhead{$p$} &
\colhead{$p_{\rm total}$} \\
\colhead{ } &
\colhead{(km)} &
\colhead{(K)} &
\colhead{(km s$^{-1}$)} &
\colhead{(cm$^{-3}$)} &
\colhead{(cm$^{-3}$)} &
\colhead{(dyn cm$^{-2}$)} &
\colhead{(dyn cm$^{-2}$)}
   }
\startdata
   1 &  19137.443 & 1586310 & 16.00 & 5.02256e08 & 6.03527e08 &  0.25318 &  0.25469 \\                                 
   2 &  14641.069 & 1442100 & 16.00 & 5.86126e08 & 7.04308e08 &  0.26860 &  0.27036 \\                                 
   3 &  11347.151 & 1311000 & 16.00 & 6.75998e08 & 8.12301e08 &  0.28162 &  0.28365 \\                                 
   4 &   8933.028 & 1191818 & 16.00 & 7.72363e08 & 9.28095e08 &  0.29252 &  0.29483 \\                                 
   5 &   7162.920 & 1083471 & 16.00 & 8.75805e08 & 1.05239e09 &  0.30154 &  0.30417 \\                                 
   6 &   5864.331 &  984974 & 16.00 & 9.87049e08 & 1.18607e09 &  0.30895 &  0.31191 \\                                 
   7 &   4910.962 &  895431 & 15.87 & 1.10713e09 & 1.33036e09 &  0.31503 &  0.31830 \\                                 
   8 &   4210.309 &  814028 & 15.72 & 1.23694e09 & 1.48634e09 &  0.31997 &  0.32355 \\                                 
   9 &   3694.627 &  740025 & 15.58 & 1.37756e09 & 1.65531e09 &  0.32395 &  0.32787 \\                                 
  10 &   3314.326 &  672750 & 15.45 & 1.53019e09 & 1.83872e09 &  0.32713 &  0.33141 \\                                 
  11 &   3033.134 &  611591 & 15.31 & 1.69619e09 & 2.03818e09 &  0.32965 &  0.33431 \\                                 
  12 &   2824.539 &  555992 & 15.18 & 1.87700e09 & 2.25545e09 &  0.33163 &  0.33670 \\                                 
  13 &   2669.179 &  505447 & 15.05 & 2.07422e09 & 2.49242e09 &  0.33316 &  0.33866 \\                                 
  14 &   2552.916 &  459497 & 14.92 & 2.28957e09 & 2.75118e09 &  0.33431 &  0.34029 \\                                
  15 &   2465.432 &  417725 & 14.79 & 2.52491e09 & 3.03396e09 &  0.33516 &  0.34164 \\                                 
  16 &   2399.193 &  379750 & 14.67 & 2.78229e09 & 3.34320e09 &  0.33575 &  0.34276 \\                                 
  17 &   2348.693 &  345227 & 14.54 & 3.06388e09 & 3.68154e09 &  0.33612 &  0.34371 \\                                 
  18 &   2309.907 &  313843 & 14.42 & 3.37209e09 & 4.05183e09 &  0.33630 &  0.34451 \\                                 
  19 &   2279.888 &  285312 & 14.29 & 3.70949e09 & 4.45719e09 &  0.33631 &  0.34519 \\                                 
  20 &   2256.477 &  259374 & 14.17 & 4.07890e09 & 4.90095e09 &  0.33618 &  0.34577 \\                                 
  21 &   2238.085 &  235795 & 14.04 & 4.48335e09 & 5.38674e09 &  0.33592 &  0.34628 \\                                 
  22 &   2223.664 &  214539 & 13.92 & 4.92206e09 & 5.91358e09 &  0.33554 &  0.34671 \\                                 
  23 &   2212.000 &  194872 & 13.80 & 5.41083e09 & 6.50032e09 &  0.33503 &  0.34710 \\                                 
  24 &   2204.472 &  180500 & 13.70 & 5.83345e09 & 7.00735e09 &  0.33454 &  0.34737 \\                                 
  25 &   2198.931 &  168800 & 13.61 & 6.22914e09 & 7.48170e09 &  0.33406 &  0.34758 \\                                 
  26 &   2194.893 &  159500 & 13.54 & 6.58401e09 & 7.90649e09 &  0.33360 &  0.34775 \\                                 
  27 &   2191.868 &  152000 & 13.48 & 6.90102e09 & 8.28539e09 &  0.33318 &  0.34788 \\                                 
  28 &   2189.411 &  145500 & 13.43 & 7.20177e09 & 8.64374e09 &  0.33278 &  0.34799 \\                                 
  29 &   2187.102 &  139000 & 13.37 & 7.52990e09 & 9.03416e09 &  0.33233 &  0.34810 \\                                 
  30 &   2183.821 &  129000 & 13.27 & 8.09855e09 & 9.70684e09 &  0.33155 &  0.34826 \\                                 
  31 &   2182.314 &  124000 & 13.22 & 8.41851e09 & 1.00793e10 &  0.33110 &  0.34834 \\                                 
  32 &   2180.806 &  118700 & 13.17 & 8.78566e09 & 1.05073e10 &  0.33058 &  0.34843 \\                                 
  33 &   2179.544 &  114000 & 13.12 & 9.14062e09 & 1.09165e10 &  0.33008 &  0.34850 \\                                 
  34 &   2178.178 &  108600 & 13.06 & 9.58523e09 & 1.14279e10 &  0.32944 &  0.34858 \\                                 
  35 &   2176.627 &  102000 & 12.97 & 1.01957e10 & 1.21162e10 &  0.32857 &  0.34868 \\                                 
  36 &   2175.203 &   95400 & 12.89 & 1.08914e10 & 1.28898e10 &  0.32758 &  0.34878 \\                                 
  37 &   2173.940 &   89000 & 12.80 & 1.16676e10 & 1.37344e10 &  0.32647 &  0.34887 \\                                 
  38 &   2172.652 &   81800 & 12.69 & 1.26853e10 & 1.48259e10 &  0.32503 &  0.34897 \\                                 
  39 &   2171.567 &   75000 & 12.58 & 1.38456e10 & 1.60013e10 &  0.32340 &  0.34906 \\                                 
  40 &   2170.518 &   67500 & 12.44 & 1.53704e10 & 1.75671e10 &  0.32128 &  0.34916 \\                                 
  41 &   2169.648 &   60170 & 12.30 & 1.71934e10 & 1.94618e10 &  0.31879 &  0.34925 \\                                 
  42 &   2168.968 &   53280 & 12.14 & 1.93269e10 & 2.16887e10 &  0.31593 &  0.34933 \\                                 
  43 &   2168.640 &   49390 & 12.05 & 2.07799e10 & 2.31927e10 &  0.31402 &  0.34937 \\                                 
  44 &   2168.338 &   45420 & 11.95 & 2.24914e10 & 2.49821e10 &  0.31181 &  0.34941 \\                                 
  45 &   2168.047 &   41180 & 11.83 & 2.46724e10 & 2.72153e10 &  0.30904 &  0.34946 \\                                 
  46 &   2167.758 &   36590 & 11.68 & 2.75621e10 & 3.01456e10 &  0.30545 &  0.34950 \\                                 
  47 &   2167.491 &   32150 & 11.52 & 3.11311e10 & 3.36001e10 &  0.30115 &  0.34956 \\                                 
  48 &   2167.237 &   27970 & 11.35 & 3.54554e10 & 3.76745e10 &  0.29610 &  0.34961 \\                                 
  49 &   2166.984 &   24060 & 11.17 & 4.07510e10 & 4.25125e10 &  0.29013 &  0.34967 \\                                 
  50 &   2166.865 &   22320 & 11.08 & 4.36563e10 & 4.50911e10 &  0.28694 &  0.34970 \\                                 
  51 &   2166.726 &   20420 & 10.97 & 4.73932e10 & 4.82202e10 &  0.28292 &  0.34974 \\                                 
  52 &   2166.628 &   19200 & 10.90 & 5.01440e10 & 5.04774e10 &  0.28003 &  0.34977 \\                                 
  53 &   2166.513 &   17930 & 10.81 & 5.34115e10 & 5.29948e10 &  0.27663 &  0.34981 \\                                 
  54 &   2166.336 &   16280 & 10.70 & 5.83302e10 & 5.66880e10 &  0.27164 &  0.34988 \\                                 
  55 &   2166.222 &   15370 & 10.63 & 6.15022e10 & 5.88678e10 &  0.26848 &  0.34992 \\                                 
  56 &   2166.101 &   14520 & 10.56 & 6.47530e10 & 6.11122e10 &  0.26530 &  0.34997 \\                                 
  57 &   2165.981 &   13800 & 10.50 & 6.79348e10 & 6.29084e10 &  0.26224 &  0.35002 \\                                 
  58 &   2165.835 &   13080 & 10.44 & 7.14196e10 & 6.48243e10 &  0.25894 &  0.35009 \\                                 
  59 &   2165.602 &   12190 & 10.35 & 7.64547e10 & 6.69811e10 &  0.25427 &  0.35020 \\                                 
  60 &   2165.328 &   11440 & 10.27 & 8.14419e10 & 6.85481e10 &  0.24977 &  0.35034 \\                                 
  61 &   2165.018 &   10850 & 10.19 & 8.63489e10 & 6.88656e10 &  0.24545 &  0.35050 \\                                 
  62 &   2164.634 &   10340 & 10.12 & 9.09834e10 & 6.90823e10 &  0.24150 &  0.35072 \\                                 
  63 &   2164.253 &    9983 & 10.07 & 9.48652e10 & 6.85288e10 &  0.23828 &  0.35095 \\                                 
  64 &   2163.862 &    9735 & 10.03 & 9.79343e10 & 6.77341e10 &  0.23583 &  0.35119 \\                                 
  65 &   2163.494 &    9587 & 10.00 & 1.00149e11 & 6.67284e10 &  0.23414 &  0.35142 \\                                 
  66 &   2163.250 &    9530 &  9.97 & 1.02083e11 & 6.45081e10 &  0.23263 &  0.35158 \\                                 
  67 &   2162.901 &    9485 &  9.95 & 1.03752e11 & 6.25925e10 &  0.23142 &  0.35181 \\                                 
  68 &   2162.500 &    9458 &  9.94 & 1.04488e11 & 6.20088e10 &  0.23106 &  0.35208 \\                                 
  69 &   2161.800 &    9425 &  9.92 & 1.06223e11 & 5.99458e10 &  0.23005 &  0.35256 \\                                 
  70 &   2159.500 &    9393 &  9.89 & 1.08782e11 & 5.72841e10 &  0.22947 &  0.35414 \\                                 
  71 &   2156.730 &    9358 &  9.87 & 1.10864e11 & 5.58102e10 &  0.22967 &  0.35610 \\                                 
  72 &   2150.001 &    9285 &  9.83 & 1.13840e11 & 5.57798e10 &  0.23204 &  0.36095 \\                                 
  73 &   2142.730 &    9228 &  9.80 & 1.16278e11 & 5.68402e10 &  0.23538 &  0.36632 \\                                 
  74 &   2110.730 &    8988 &  9.68 & 1.28080e11 & 6.12395e10 &  0.25083 &  0.39139 \\                                 
  75 &   2062.732 &    8635 &  9.47 & 1.50748e11 & 6.55879e10 &  0.27589 &  0.43417 \\                                 
  76 &   2008.731 &    8273 &  9.21 & 1.83679e11 & 6.87908e10 &  0.30936 &  0.49191 \\                                 
  77 &   1952.734 &    7970 &  8.93 & 2.27003e11 & 7.12501e10 &  0.35317 &  0.56545 \\                                 
  78 &   1903.909 &    7780 &  8.69 & 2.73100e11 & 7.37062e10 &  0.40186 &  0.64358 \\                                 
  79 &   1841.845 &    7600 &  8.38 & 3.46586e11 & 7.73687e10 &  0.48122 &  0.76636 \\                                 
  80 &   1761.577 &    7410 &  7.96 & 4.78227e11 & 8.18332e10 &  0.62191 &  0.97675 \\                                 
  81 &   1667.773 &    7220 &  7.44 & 7.12784e11 & 8.64956e10 &  0.86780 &   1.3300 \\                                 
  82 &   1581.856 &    7080 &  6.94 & 1.04860e12 & 9.19885e10 &   1.2174 &   1.8088 \\                                 
  83 &   1477.185 &    6910 &  6.28 & 1.73736e12 & 1.00355e11 &   1.9190 &   2.7224 \\                                 
  84 &   1369.999 &    6740 &  5.55 & 3.04564e12 & 1.11225e11 &   3.2211 &   4.3198 \\                                 
  85 &   1275.000 &    6570 &  4.84 & 5.24270e12 & 1.19466e11 &   5.3395 &   6.7774 \\                                 
  86 &   1175.000 &    6370 &  4.03 & 9.77672e12 & 1.25084e11 &   9.5683 &   11.428 \\                                 
  87 &   1080.000 &    6180 &  3.20 & 1.85649e13 & 1.30792e11 &   17.536 &   19.759 \\                                 
  88 &    985.000 &    5950 &  2.52 & 3.66227e13 & 1.27074e11 &   33.198 &   35.925 \\                                 
  89 &    915.000 &    5760 &  2.21 & 6.15603e13 & 1.18778e11 &   53.946 &   57.461 \\                                 
  90 &    855.000 &    5570 &  2.00 & 9.80298e13 & 1.06495e11 &   83.008 &   87.606 \\                                 
  91 &    805.000 &    5380 &  1.77 & 1.48029e14 & 9.24561e10 &   121.02 &   126.47 \\                                 
  92 &    755.000 &    5160 &  1.53 & 2.29150e14 & 7.84023e10 &   179.63 &   185.92 \\                                 
  93 &    705.000 &    4900 &  1.31 & 3.65711e14 & 7.04611e10 &   272.20 &   279.50 \\                                 
  94 &    650.000 &    4680 &  1.07 & 6.20897e14 & 8.73902e10 &   441.36 &   449.64 \\                                 
  95 &    600.000 &    4560 &  0.88 & 1.00588e15 & 1.27406e11 &   696.67 &   705.86 \\                                 
  96 &    560.000 &    4520 &  0.75 & 1.47211e15 & 1.78726e11 &   1010.6 &   1020.2 \\                                 
  97 &    525.000 &    4500 &  0.64 & 2.05297e15 & 2.41649e11 &   1403.1 &   1412.8 \\                                 
  98 &    490.000 &    4510 &  0.55 & 2.84463e15 & 3.27490e11 &   1948.4 &   1958.4 \\                                 
  99 &    450.000 &    4540 &  0.47 & 4.10533e15 & 4.63696e11 &   2830.5 &   2841.3 \\                                 
 100 &    400.000 &    4610 &  0.39 & 6.41797e15 & 7.16110e11 &   4493.0 &   4504.7 \\                                 
 101 &    350.000 &    4690 &  0.34 & 9.93753e15 & 1.09932e12 &   7077.3 &   7090.7 \\                                 
 102 &    300.000 &    4780 &  0.36 & 1.52241e16 & 1.67943e12 &    11050 &    11072 \\                                 
 103 &    250.000 &    4880 &  0.48 & 2.30399e16 & 2.55667e12 &    17071 &    17134 \\                                 
 104 &    200.000 &    4990 &  0.67 & 3.43977e16 & 3.87993e12 &    26059 &    26241 \\                                 
 105 &    175.000 &    5060 &  0.74 & 4.17477e16 & 4.80290e12 &    32070 &    32335 \\                                 
 106 &    150.000 &    5150 &  0.88 & 5.01933e16 & 5.98466e12 &    39244 &    39698 \\                                 
 107 &    125.000 &    5270 &  0.99 & 5.97639e16 & 7.58537e12 &    47817 &    48507 \\                                 
 108 &    100.000 &    5410 &  1.13 & 7.04752e16 & 9.81795e12 &    57888 &    58943 \\                                 
 109 &     75.000 &    5580 &  1.26 & 8.22144e16 & 1.32977e13 &    69658 &    71183 \\                                 
 110 &     50.000 &    5790 &  1.41 & 9.45666e16 & 1.95111e13 &    83148 &    85363 \\                                 
 111 &     35.000 &    5980 &  1.49 & 1.01469e17 & 2.74972e13 &    92155 &    94806 \\                                 
 112 &     20.000 &    6180 &  1.57 & 1.08418e17 & 4.02914e13 &   101773 &   104911 \\                                 
 113 &     10.000 &    6340 &  1.63 & 1.12665e17 & 5.45801e13 &   108512 &   112009 \\                                 
 114 &      0.000 &    6520 &  1.67 & 1.16638e17 & 7.64303e13 &   115549 &   119371 \\                                 
 115 &    -10.000 &    6720 &  1.71 & 1.20298e17 & 1.10009e14 &   122862 &   126978 \\                                 
 116 &    -20.000 &    6980 &  1.76 & 1.22821e17 & 1.71976e14 &   130354 &   134788 \\                                 
 117 &    -30.000 &    7280 &  1.79 & 1.24613e17 & 2.78958e14 &   138048 &   142734 \\                                 
 118 &    -40.000 &    7590 &  1.86 & 1.25982e17 & 4.44610e14 &   145679 &   150779 \\                                 
 119 &    -50.000 &    7900 &  1.89 & 1.27397e17 & 6.87037e14 &   153594 &   158914 \\                                 
 120 &    -60.000 &    8220 &  1.94 & 1.28397e17 & 1.04128e15 &   161466 &   167126 \\                                 
 121 &    -70.000 &    8540 &  2.00 & 1.29174e17 & 1.53163e15 &   169340 &   175393 \\                                 
 122 &    -80.000 &    8860 &  2.00 & 1.30008e17 & 2.19433e15 &   177618 &   183710 \\                                 
 123 &    -90.000 &    9140 &  2.00 & 1.31273e17 & 2.95244e15 &   185945 &   192096 \\                                 
 124 &   -100.000 &    9400 &  2.00 & 1.32657e17 & 3.83198e15 &   194352 &   200569 \\                                 
\enddata
\end{deluxetable}


\clearpage

\begin{deluxetable}{ccccc}
\tablecaption{ {\sc The Logarithmic Gradients of $T$ and $n_{\rm p} / n_{\rm a}$
(in cm$^{-1}$) at $T = 2 \times 10^4$ and at $T = 10^5$ K}  \label{TAB2} }
\tablewidth{0pt}
\tablehead{
\colhead{{\sc Model}} &
\colhead{ $d \log T / dz$} & 
\colhead{ $d \log (n_{\rm p} / n_{\rm a}) / dz$ } &
\colhead{ $d \log T / dz$} &
\colhead{ $d \log (n_{\rm p} / n_{\rm a}) / dz$ } \\
\colhead{ } &
\colhead{at $T = 2 \times 10^4$ K} &
\colhead{at $T = 2 \times 10^4$ K} &
\colhead{at $10^5$ K} &
\colhead{at $10^5$ K}
}
\startdata
in10 & $4.99 \times 10^{-7}$ & $2.73 \times 10^{-6}$ &
       $8.12 \times 10^{-8}$ & $1.98 \times 10^{-7}$   \\
in5  & $1.12 \times 10^{-6}$ & $5.91 \times 10^{-6}$ &
       $1.58 \times 10^{-7}$ & $3.86 \times 10^{-7}$   \\
in1  & $4.48 \times 10^{-6}$ & $1.09 \times 10^{-5}$ &
       $3.24 \times 10^{-7}$ & $8.00 \times 10^{-7}$   \\
0    & $6.26 \times 10^{-6}$ & $1.05 \times 10^{-5}$ &
       $4.59 \times 10^{-7}$ & $1.24 \times 10^{-6}$   \\
out1 & $7.59 \times 10^{-6}$ & $1.13 \times 10^{-5}$ &
       $6.70 \times 10^{-7}$ & $3.79 \times 10^{-6}$   \\
out2 & $9.67 \times 10^{-6}$ & $1.13 \times 10^{-5}$ &
       $8.74 \times 10^{-7}$ & $9.65 \times 10^{-6}$ 
\enddata
\end{deluxetable}


\clearpage

\begin{deluxetable}{ccccccc}
\rotate
\tablecaption{ {\sc The Heat Flux and Its Components at $T = 2 \times 10^4$ K.}
   \label{TAB3} }
\tablewidth{0pt}
\tablehead{ \colhead{{\sc Model}} &
\colhead{$F_{\rm H}$} &
\colhead{$z-z_0$} &
\colhead{$F_h$} &
\colhead{$F_{\rm cond}$} &
\colhead{$F_{U{\rm react}}$} &
\colhead{$F_{\rm react.total}$} \\
\colhead{ } &
\colhead{(cm$^{-2}$ s$^{-1}$)} & 
\colhead{(cm)} &
\colhead{(erg cm$^{-2}$ s$^{-1}$)} &
\colhead{(erg cm$^{-2}$ s$^{-1}$)} &
\colhead{(erg cm$^{-2}$ s$^{-1}$)} &
\colhead{(erg cm$^{-2}$ s$^{-1}$)} 
}
\startdata
in10   & $-10 \times 10^{15}$ & $2.98 \times 10^6$ & $-1.28 \times 10^5$ &
         $-9.44 \times 10^2$ & $2.91 \times 10^4$ & $2.85 \times 10^4$ \\
in5    & $ -5 \times 10^{15}$ & $1.47 \times 10^6$ & $-6.70 \times 10^4$ &
         $-2.14 \times 10^3$ & $1.52 \times 10^4$ & $1.33 \times 10^4$ \\
in1    & $ -1 \times 10^{15}$ & $4.94 \times 10^5$ & $-3.45 \times 10^4$ &
         $-8.87 \times 10^3$ & $4.56 \times 10^3$ & $-6.91 \times 10^3$ \\
0      & $0$                  & $3.44 \times 10^5$ & $-3.33 \times 10^4$ &
         $-1.23 \times 10^4$ & $0$                 & $-1.60 \times 10^4$ \\
out1   & $ +1 \times 10^{15}$ & $2.53 \times 10^5$ & $-4.27 \times 10^4$ &
         $-1.52 \times 10^4$ & $-1.30 \times 10^4$ & $-3.48 \times 10^4$ \\
out2   & $ +2 \times 10^{15}$ & $2.04 \times 10^5$ & $-5.42 \times 10^4$ &
         $-1.93 \times 10^4$ & $-2.63 \times 10^4$ & $-5.42 \times 10^4$
\enddata
\end{deluxetable}


\clearpage

\begin{deluxetable}{ccccccc}
\rotate
\tablecaption{ {\sc The Heat Flux and Its Components at $T = 10^5$ K.}
   \label{TAB4} }
\tablewidth{0pt}
\tablehead{ \colhead{{\sc Model}} &
\colhead{$F_{\rm H}$} &
\colhead{$z-z_0$} &
\colhead{$F_h$} &
\colhead{$F_{\rm cond}$} &
\colhead{$F_{U{\rm react}}$} &
\colhead{$F_{\rm react.total}$} \\
\colhead{ } &
\colhead{(cm$^{-2}$ s$^{-1}$)} & 
\colhead{(cm)} &
\colhead{(erg cm$^{-2}$ s$^{-1}$)} &
\colhead{(erg cm$^{-2}$ s$^{-1}$)} &
\colhead{(erg cm$^{-2}$ s$^{-1}$)} &
\colhead{(erg cm$^{-2}$ s$^{-1}$)} 
}
\startdata
in10   & $-10 \times 10^{15}$ & $1.05 \times 10^7$ & $-8.23 \times 10^5$ &
         $-3.19 \times 10^4$ & $1.65 \times 10^3$ & $1.65 \times 10^3$ \\
in5    & $ -5 \times 10^{15}$ & $5.42 \times 10^6$ & $-4.58 \times 10^5$ &
         $-6.23 \times 10^4$ & $8.96 \times 10^2$ & $8.89 \times 10^2$ \\
in1    & $ -1 \times 10^{15}$ & $1.92 \times 10^6$ & $-2.07 \times 10^5$ &
         $-1.28 \times 10^5$ & $2.86 \times 10^2$ & $2.56 \times 10^2$ \\
0      & $0$                  & $1.30 \times 10^6$ & $-1.81 \times 10^5$ &
         $-1.81 \times 10^5$ & $0$                 & $2.75 \times 10^2$ \\
out1   & $ +1 \times 10^{15}$ & $9.19 \times 10^5$ & $-1.96 \times 10^5$ &
         $-2.64 \times 10^5$ & $-7.70 \times 10^3$ & $-7.94 \times 10^3$ \\
out2   & $ +2 \times 10^{15}$ & $7.17 \times 10^5$ & $-2.09 \times 10^5$ &
         $-3.44 \times 10^5$ & $-1.67 \times 10^4$ & $-1.68 \times 10^4$
\enddata
\end{deluxetable}



\clearpage

\begin{figure}
\epsscale{0.45}
\plotone{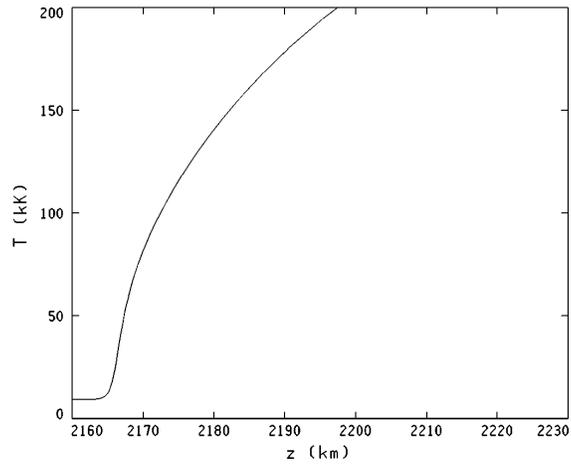}
\caption{ The temperature distribution used to determine the
results shown in Figures 2 - 7 (modified version of model C from Fontenla
et al. 1999). }
\end{figure}

\clearpage

\begin{figure}
\epsscale{0.93}
\plotone{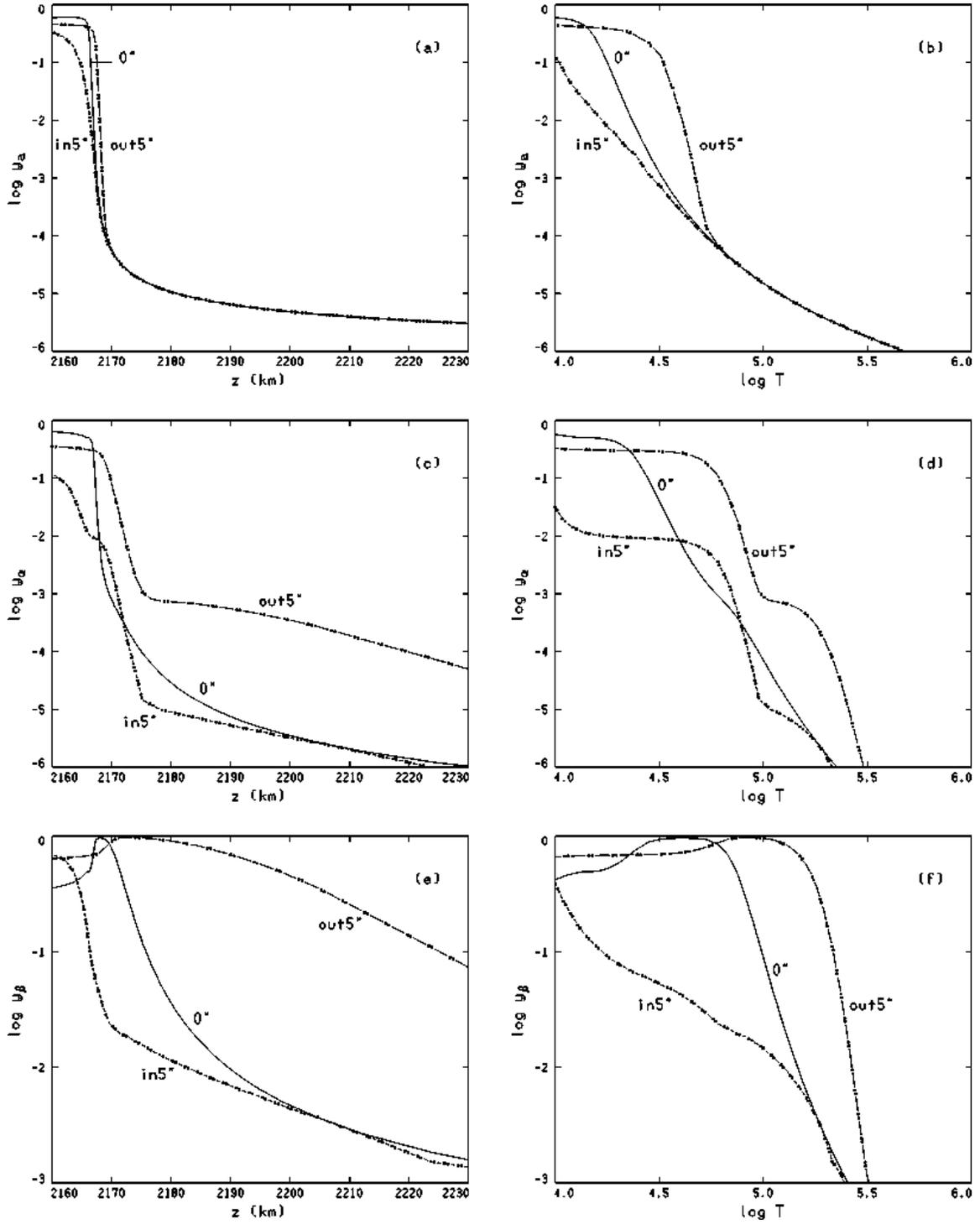}
\caption{ $y_{\rm a}$, $y_\alpha$, and $y_\beta$ vs. $z$ and
vs. $T$ for models in5$^{\prime\prime}$, 0$^{\prime\prime}$, and
out5$^{\prime\prime}$ without diffusion. }
\end{figure}

\clearpage

\begin{figure}
\epsscale{0.93}
\plotone{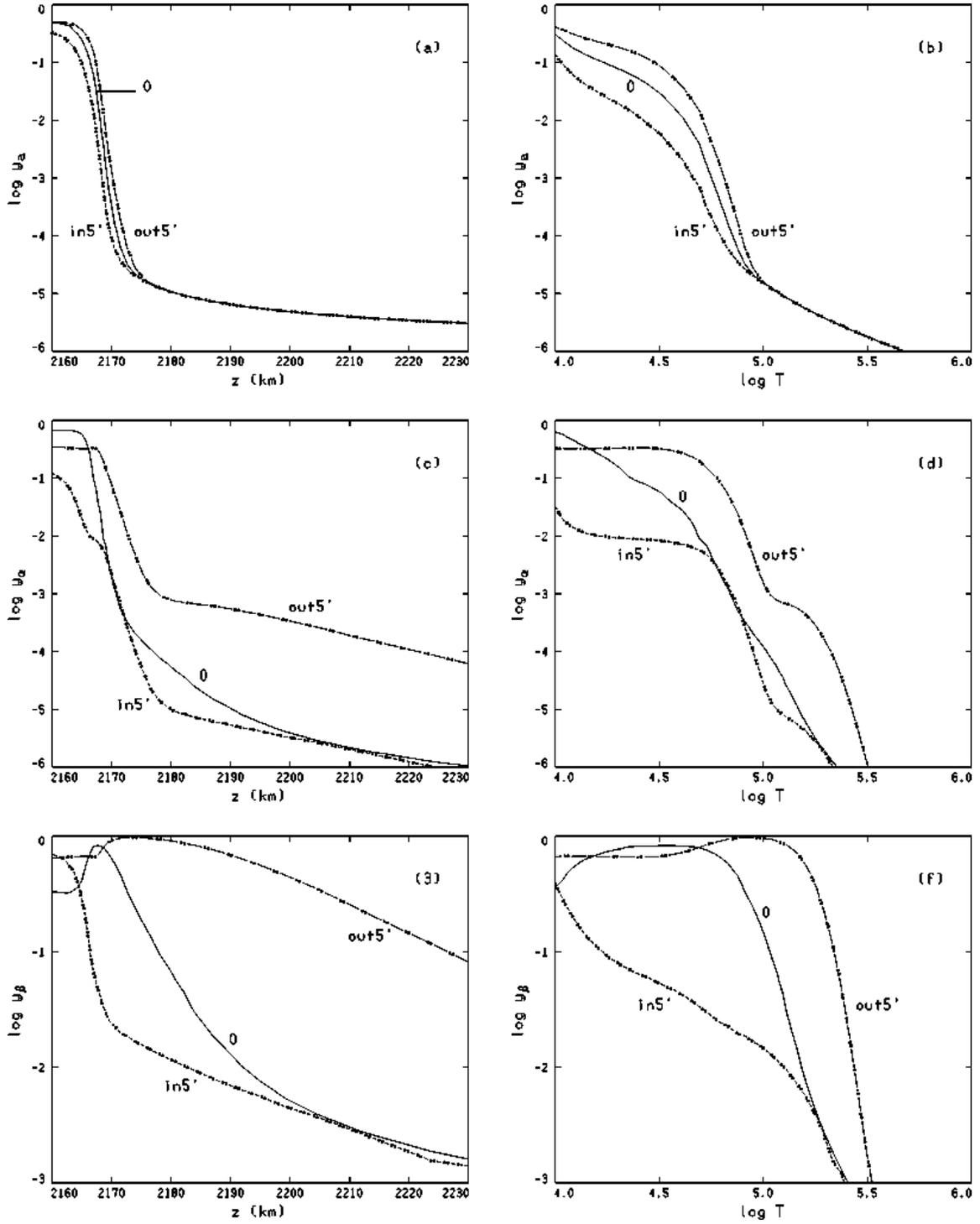}
\caption{ $y_{\rm a}$, $y_\alpha$, and $y_\beta$ vs. $z$ and
vs. $T$ for models in5$^\prime$, 0, and out5$^\prime$ that include
diffusion. }
\end{figure}

\clearpage

\begin{figure}
\epsscale{0.45}
\plotone{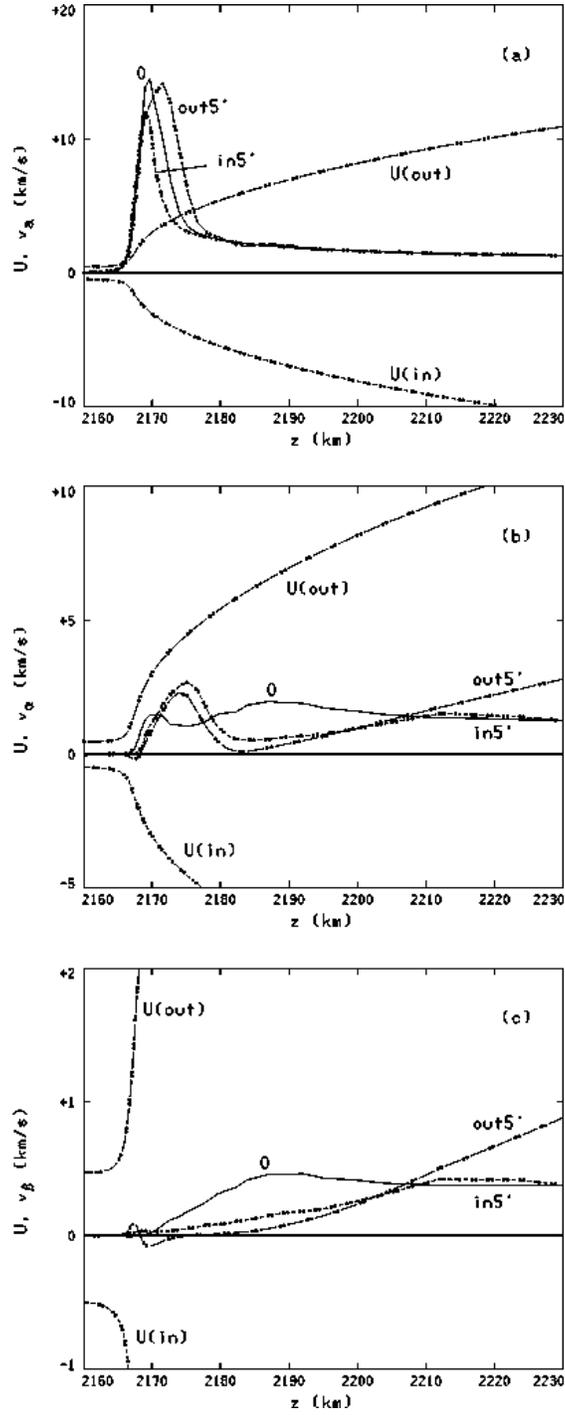}
\caption{ The flow velocity $U$, together with the H, \ion{He}{1}, and
\ion{He}{2} diffusion velocities $v_{\rm a}$, $v_\alpha$, and $v_\beta$,
respectively, for models in5$^\prime$, 0, and out5$^\prime$ that include
diffusion. }
\end{figure}

\clearpage

\begin{figure}
\epsscale{0.93}
\plotone{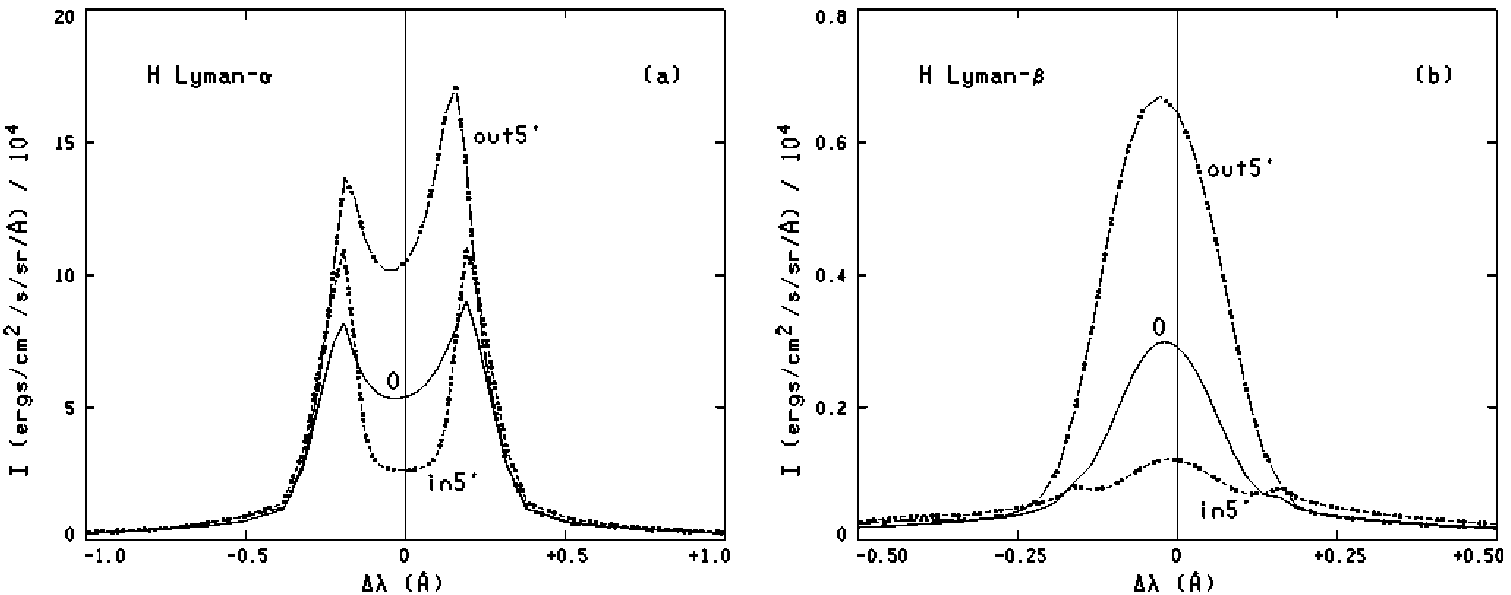}
\caption{ The calculated disk-center H line profiles for models
in5$^\prime$, 0, and out5$^\prime$. }
\end{figure}

\clearpage

\begin{figure}
\epsscale{0.45}
\plotone{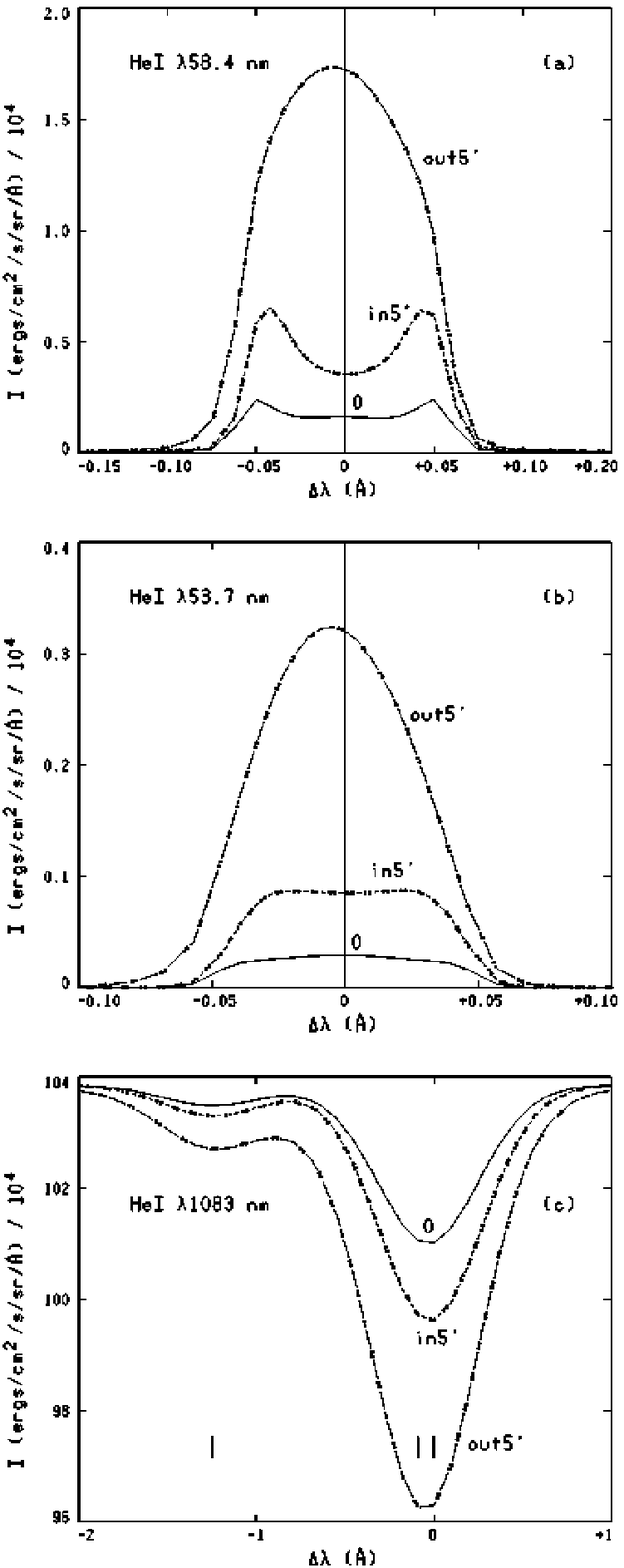}
\caption{ The calculated disk-center \ion{He}{1} line profiles
for models in5$^\prime$, 0, and out5$^\prime$. }
\end{figure}

\clearpage

\begin{figure}
\epsscale{0.45}
\plotone{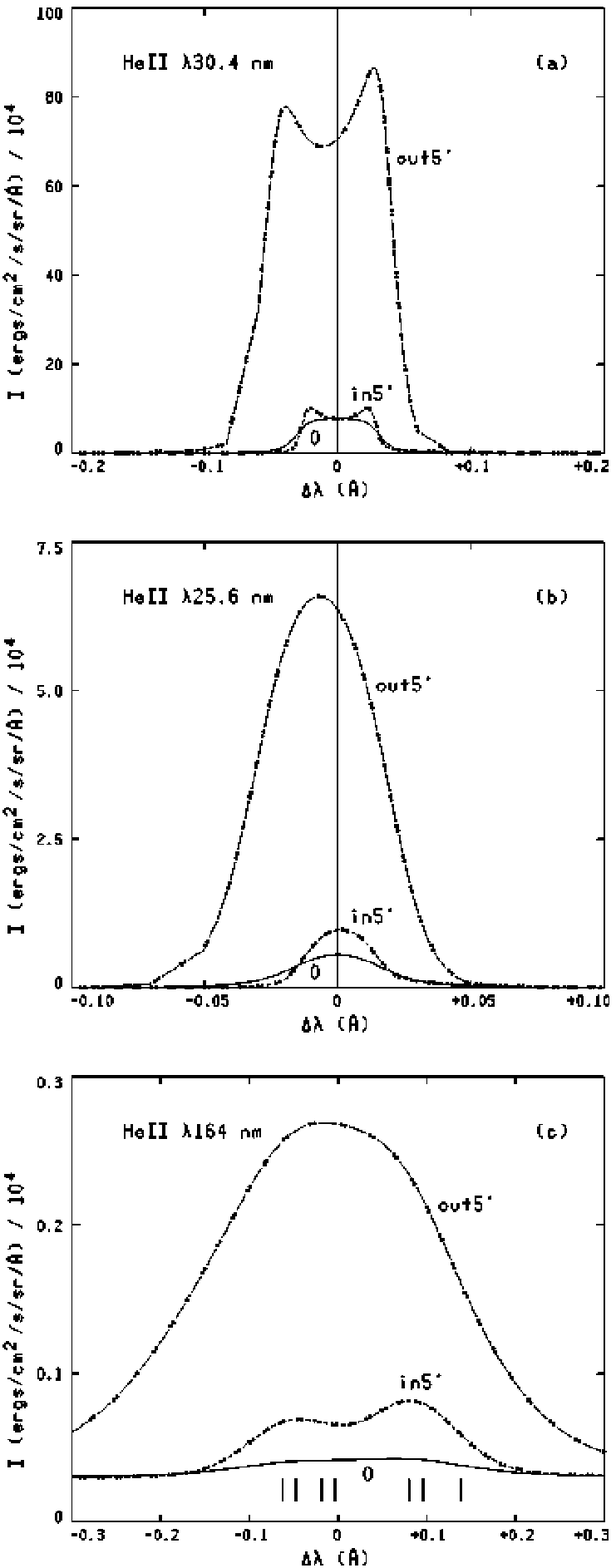}
\caption{ The calculated disk-center \ion{He}{2} line profiles
for models in5$^\prime$, 0, and out5$^\prime$. }
\end{figure}

\clearpage

\begin{figure}
\epsscale{0.93}
\plotone{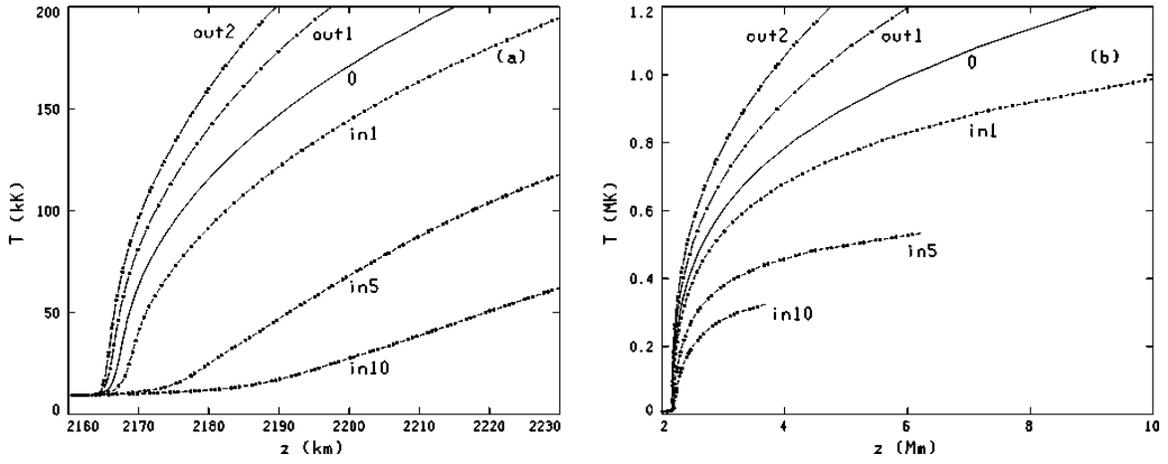}
\caption{ The calculated temperature distributions (a) in the lower
transition region, and (b) extending into the corona, for the six
energy-balance models out2, out1, 0, in1, in5, and in10. The upper
temperature limits are chosen to keep the flow velocities subsonic. }
\end{figure}

\clearpage

\begin{figure}
\epsscale{0.93}
\plotone{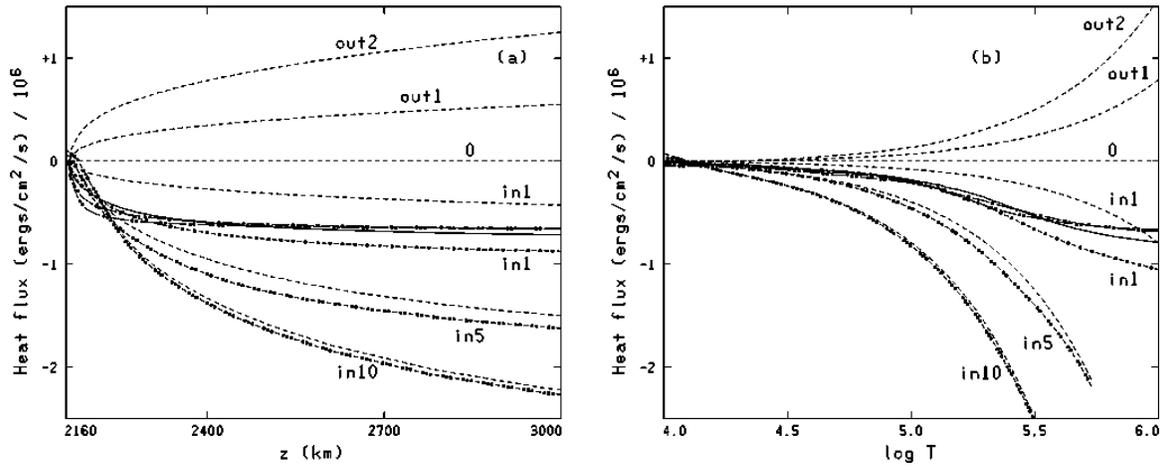}
\caption{ The total heat flux $F_h$ (curves with dots and short and
long dashes as in Fig. 8) and its velocity-driven component $F_U$
(dashed curves), from equation (48), vs. $z$ and vs. $T$. }
\end{figure}

\clearpage

\begin{figure}
\epsscale{0.93}
\plotone{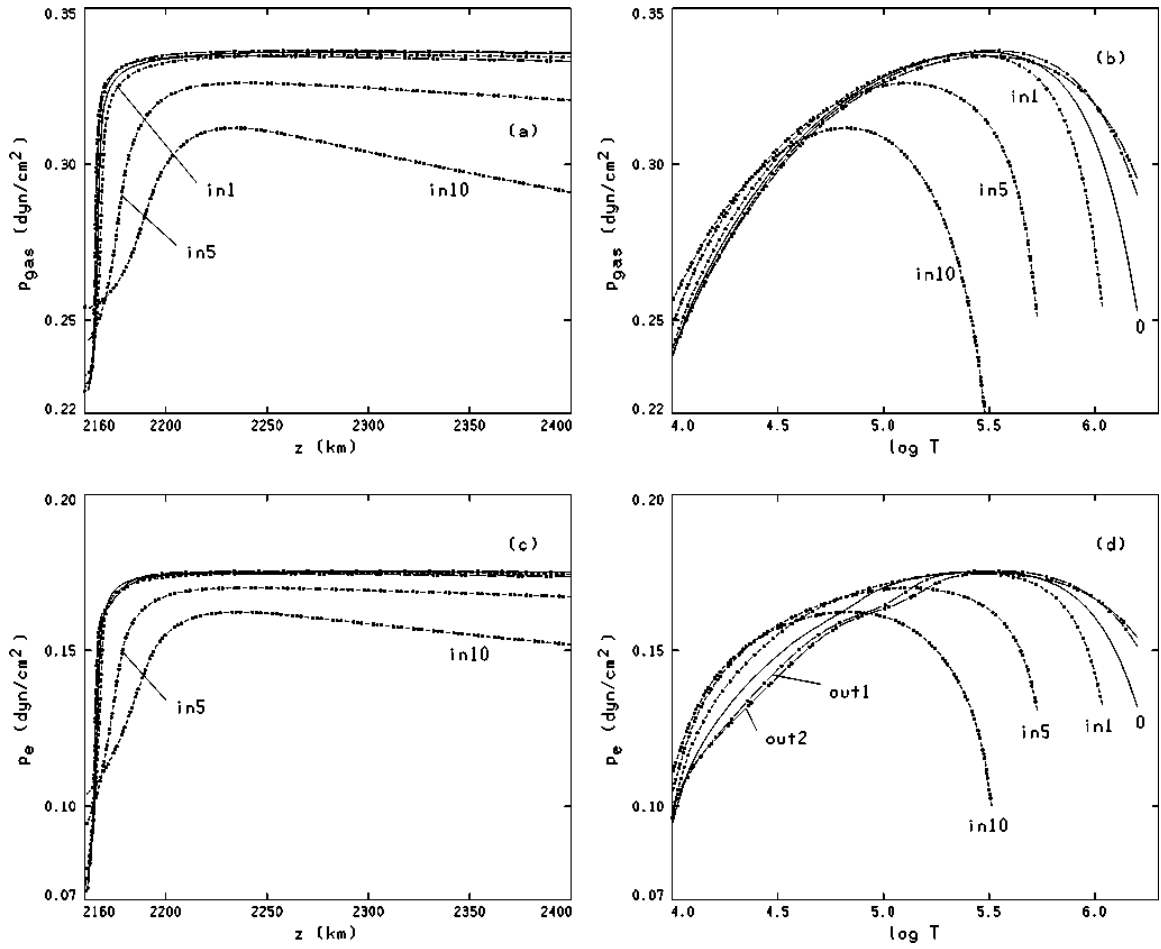}
\caption{ $p_{\rm gas}$ and $p_e$ vs. $z$ and vs. $T$. }
\end{figure}

\clearpage

\begin{figure}
\epsscale{0.45}
\plotone{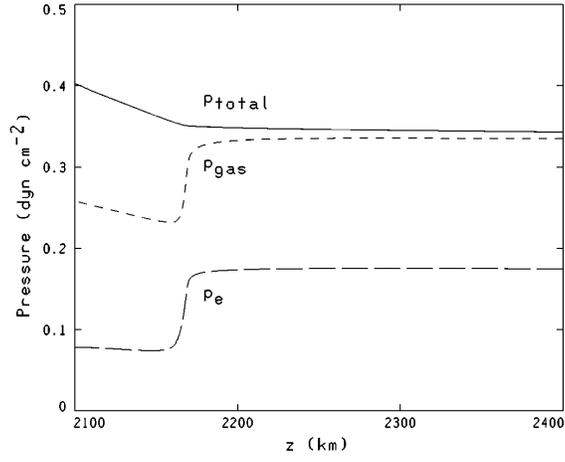}
\caption{ $p_e$, $p_{\rm gas}$ and $p_{\rm total}$ vs.
$z$ for model in1. }
\end{figure}

\clearpage

\begin{figure}
\epsscale{0.93}
\plotone{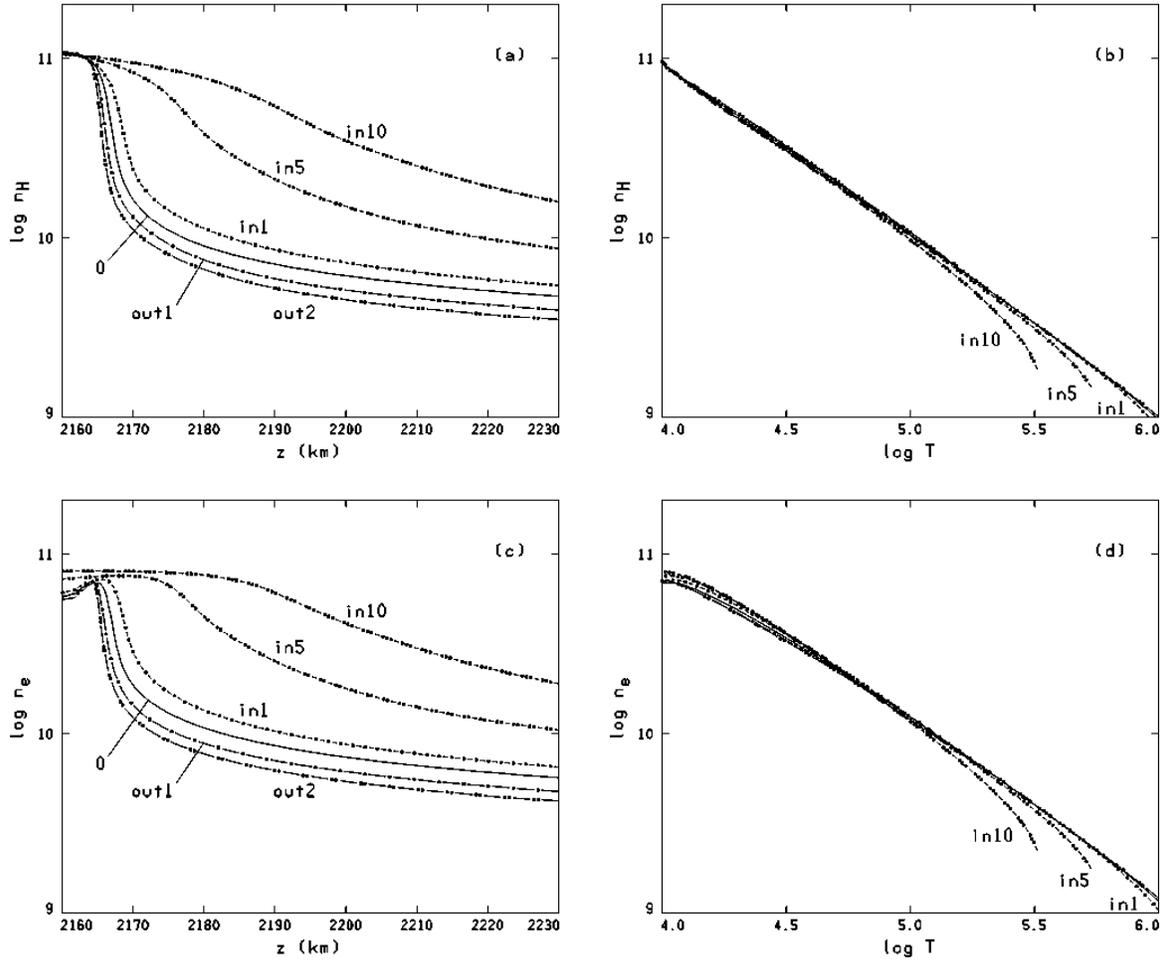}
\caption{ $n_{\rm H}$ and $n_e$ vs. $z$ and vs. $T$. }
\end{figure}

\clearpage

\begin{figure}
\epsscale{0.93}
\plotone{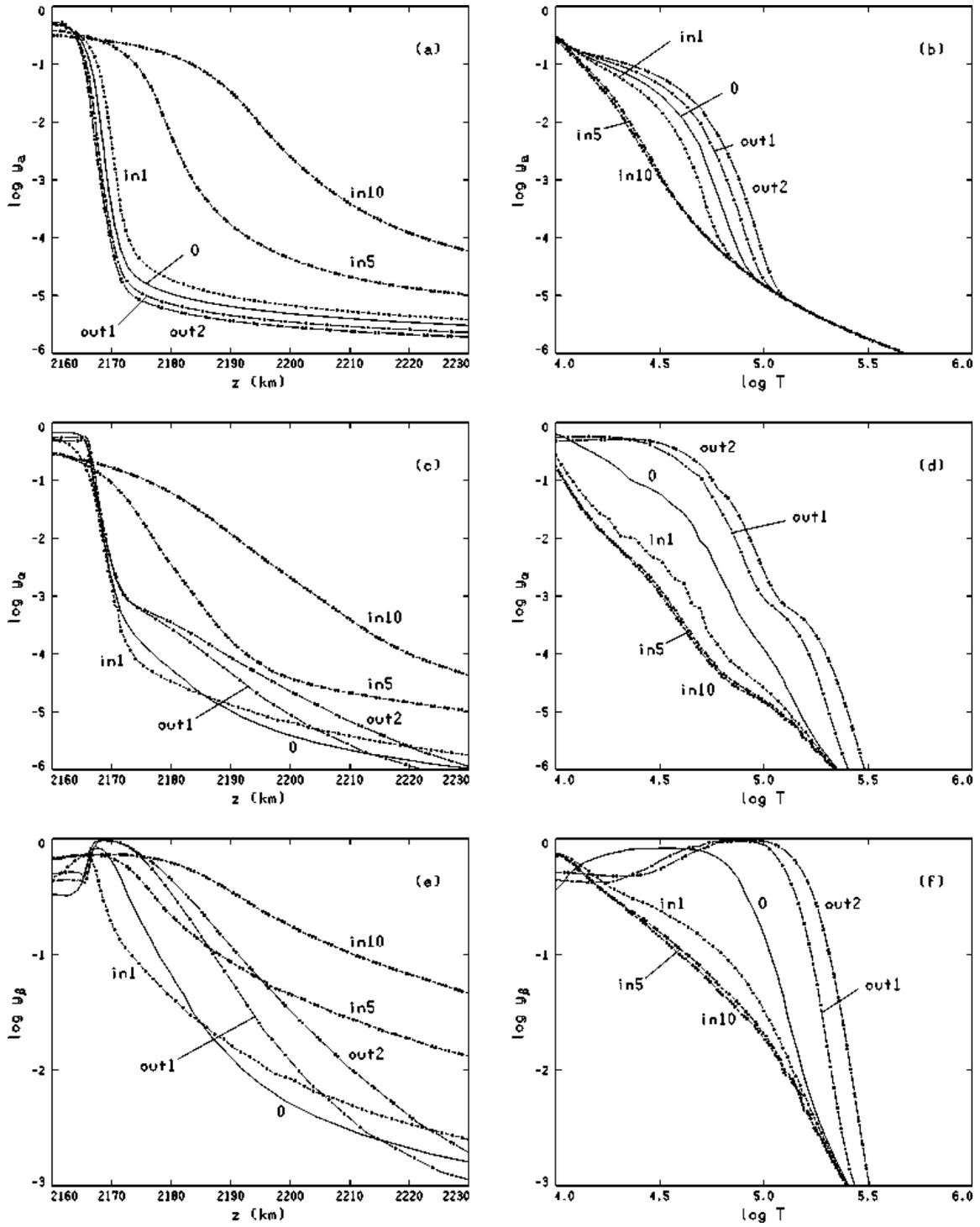}
\caption{ $y_{\rm a}$, $y_\alpha$, and $y_\beta$ vs. $z$ and
vs. $T$. }
\end{figure}

\clearpage

\begin{figure}
\epsscale{0.93}
\plotone{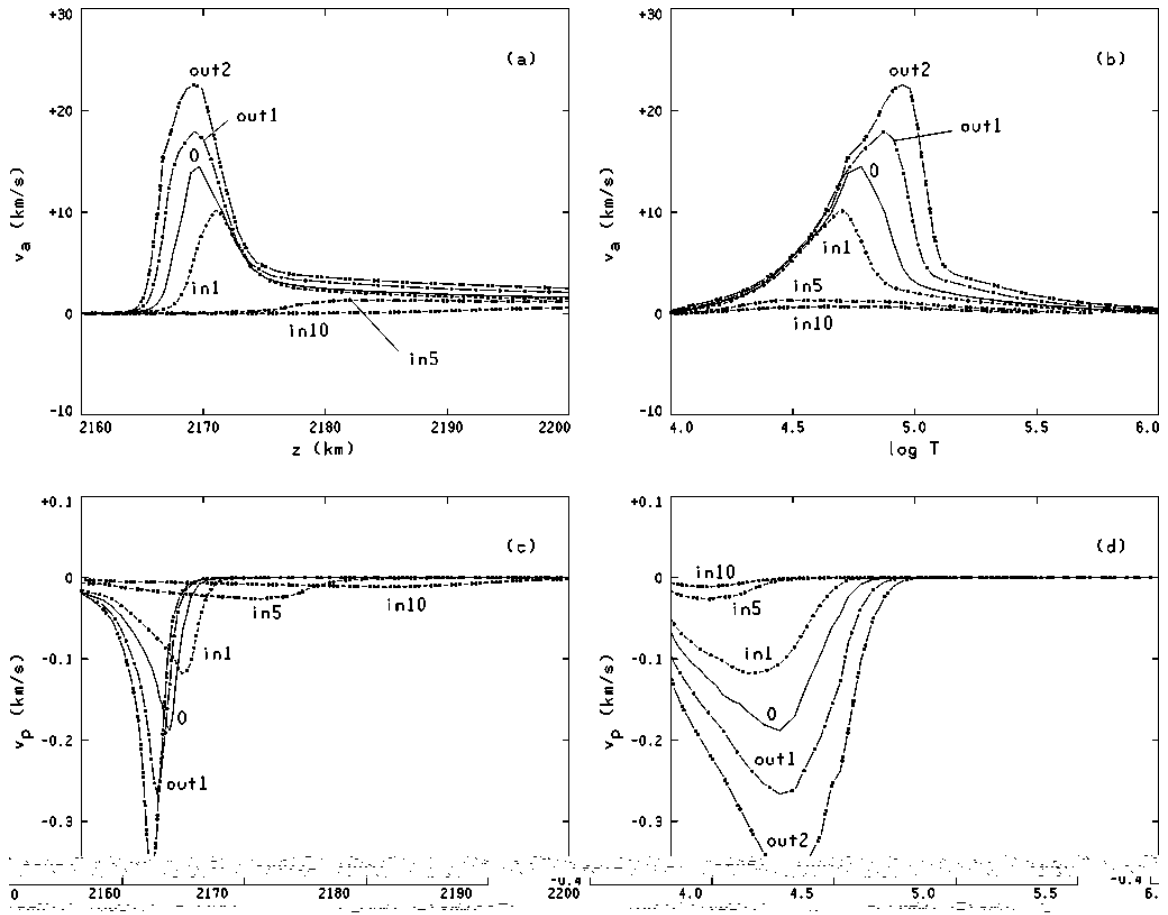}
\caption{ $v_{\rm a}$ and $v_{\rm p}$ vs. $z$ and vs. $T$. }
\end{figure}

\clearpage

\begin{figure}
\epsscale{0.93}
\plotone{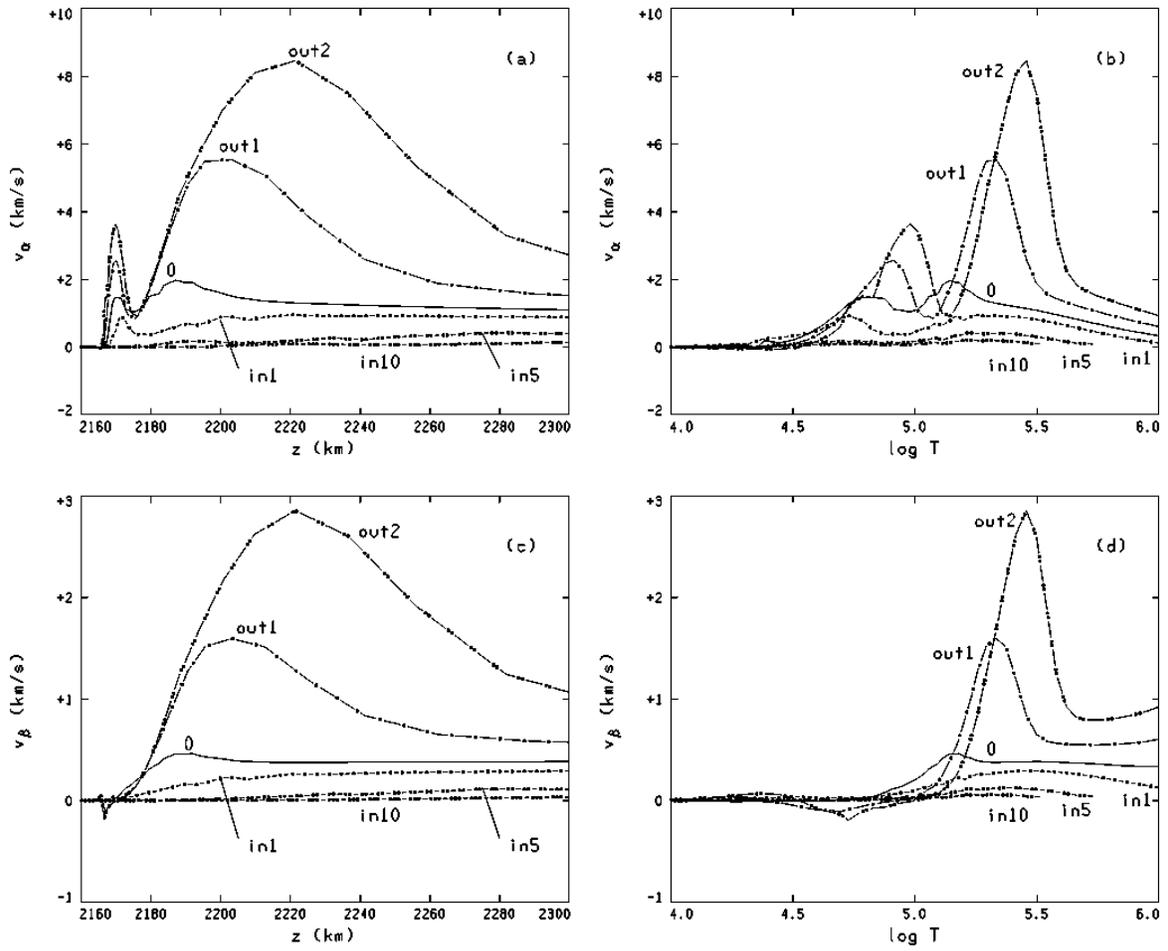}
\caption{ $v_\alpha$ and $v_\beta$ vs. $z$ and vs. $T$. }
\end{figure}

\clearpage

\begin{figure}
\epsscale{0.93}
\plotone{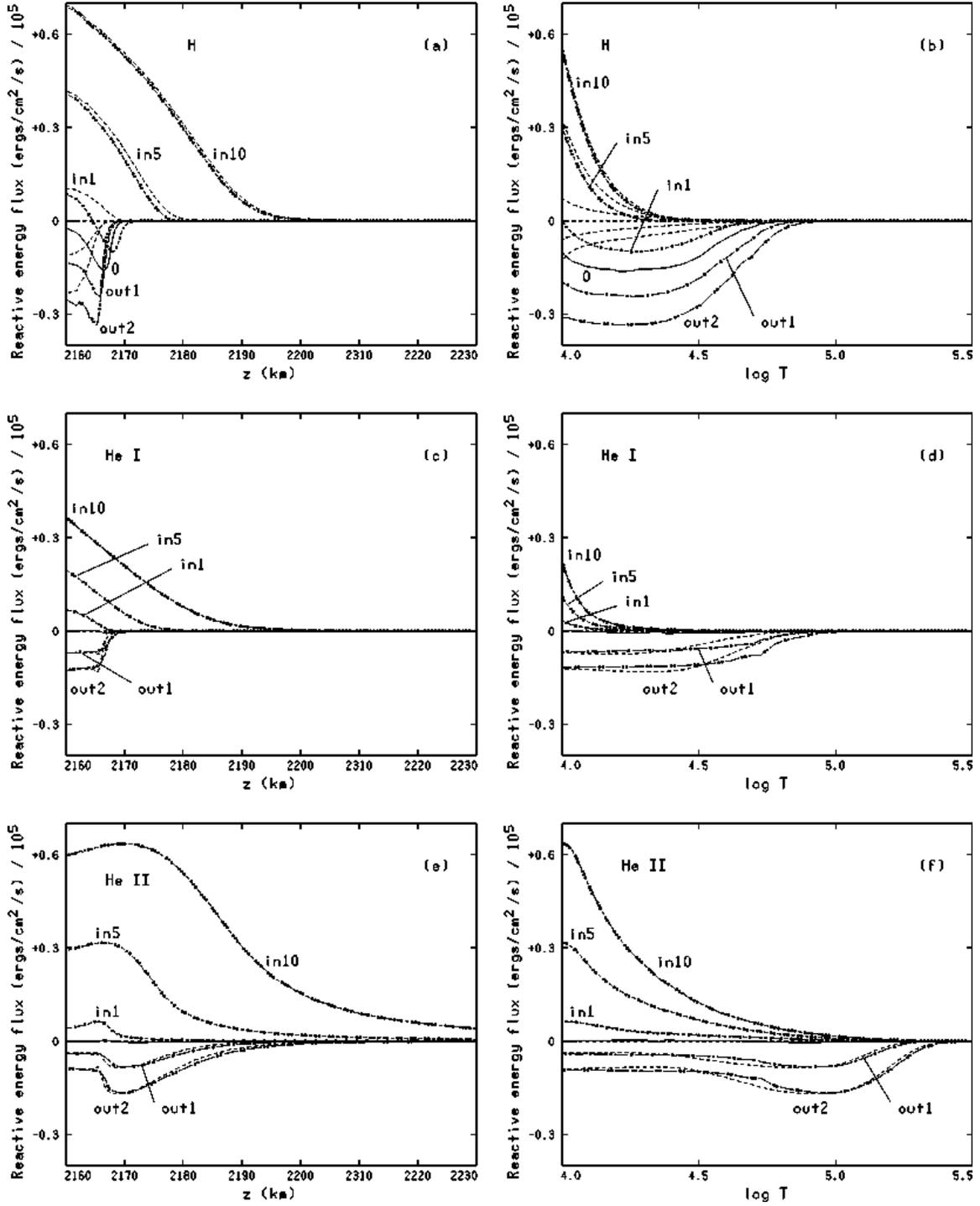}
\caption{ The H, \ion{He}{1}, and \ion{He}{2} components of
$F_{U {\rm react}}$ (dashed curves) and $F_{\rm react.total}$
(curves with dots and short and long dashes as in Fig. 8),
vs. $z$ and vs. $T$. See equations (50), (52), and (56). }
\end{figure}

\clearpage

\begin{figure}
\epsscale{0.93}
\plotone{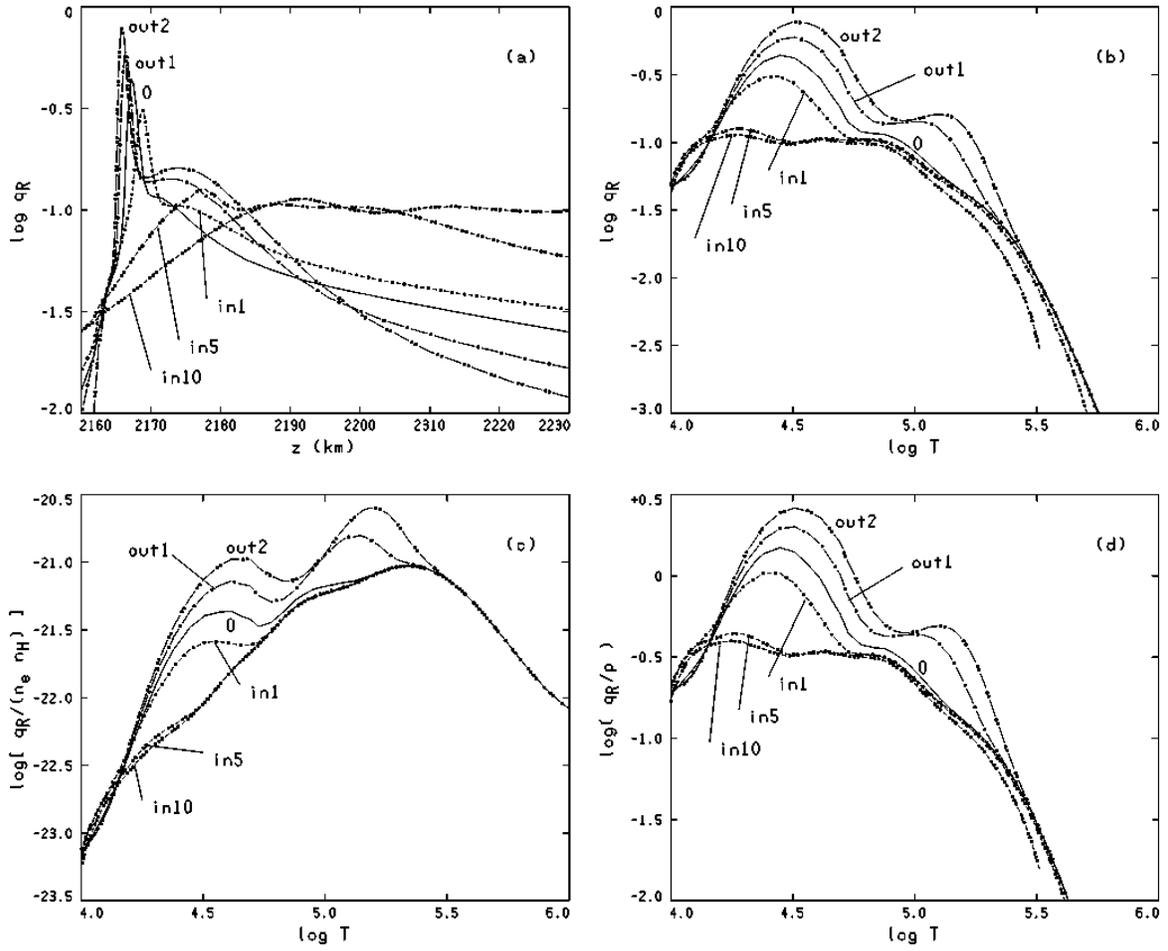}
\caption{ $q_R$ vs. $z$ and vs. $T$; $q_R / { n_{\rm e}
n_{\rm H} }$ vs. $T$; and $q_R / p$ vs. $T$. }
\end{figure}

\clearpage

\begin{figure}
\epsscale{0.93}
\plotone{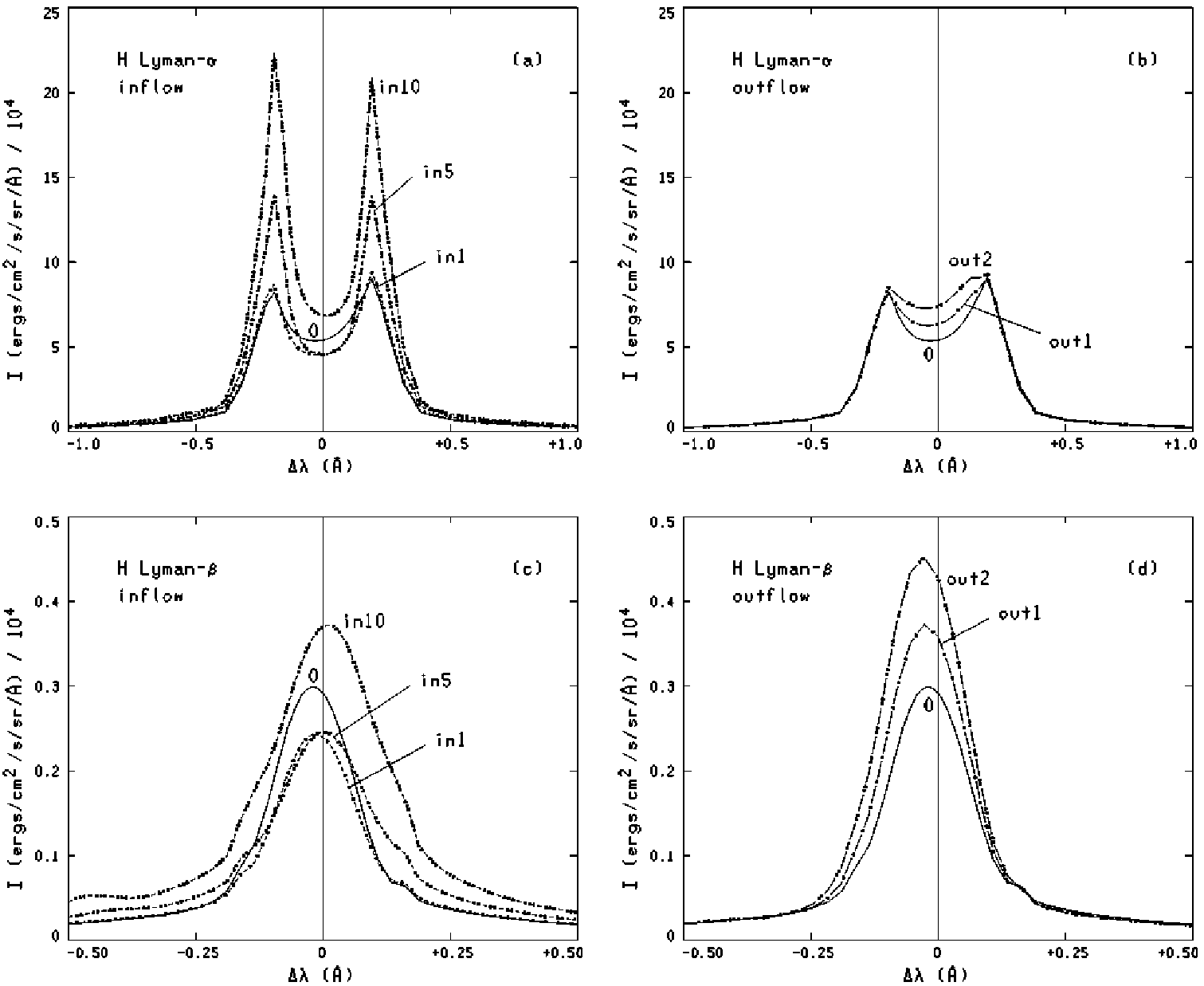}
\caption{ The calculated disk-center H line profiles for the
six energy-balance models. }
\end{figure}

\clearpage

\begin{figure}
\epsscale{0.93}
\plotone{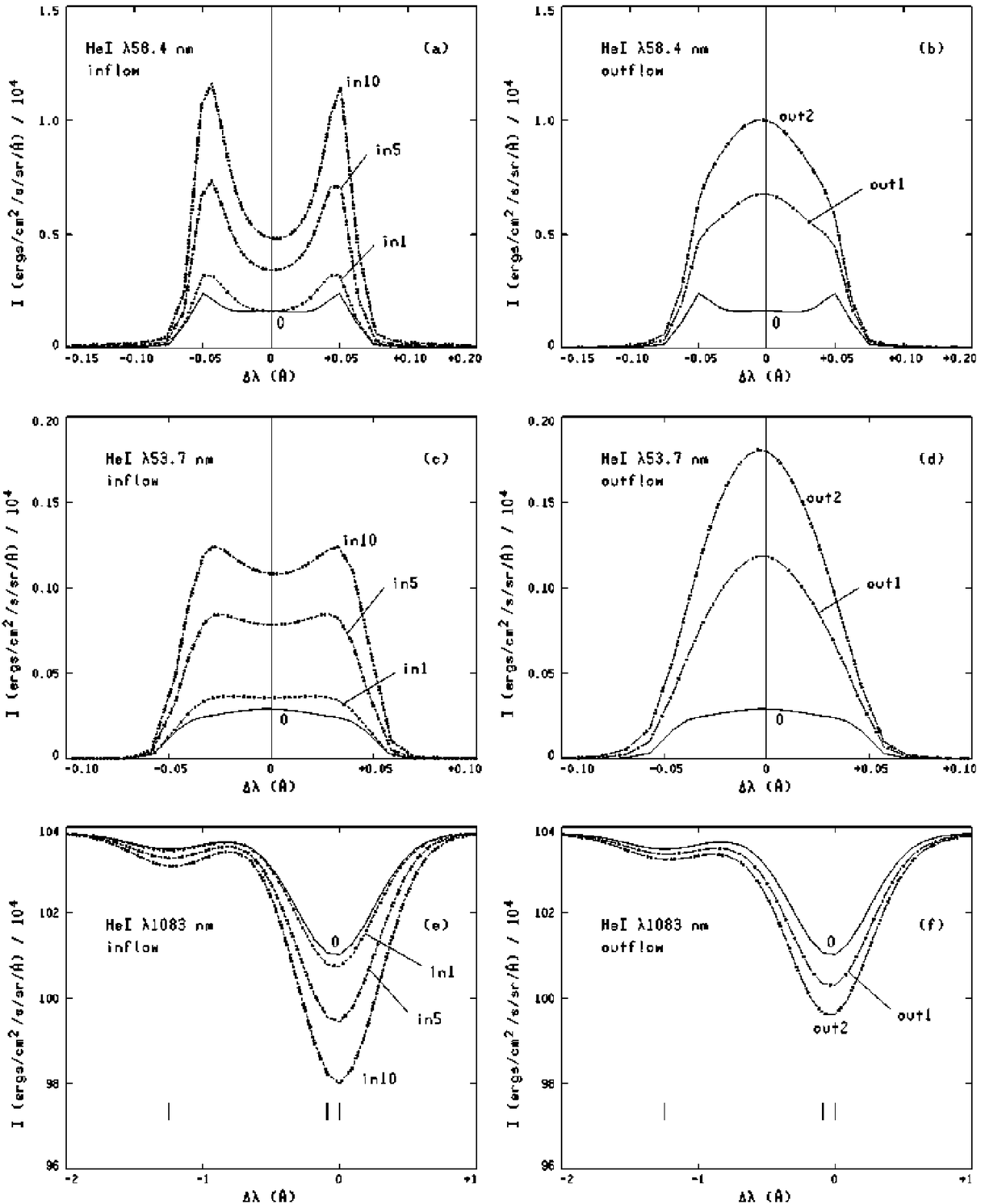}
\caption{ The calculated disk-center \ion{He}{1} line profiles
for the six energy-balance models. }
\end{figure}

\clearpage

\begin{figure}
\epsscale{0.93}
\plotone{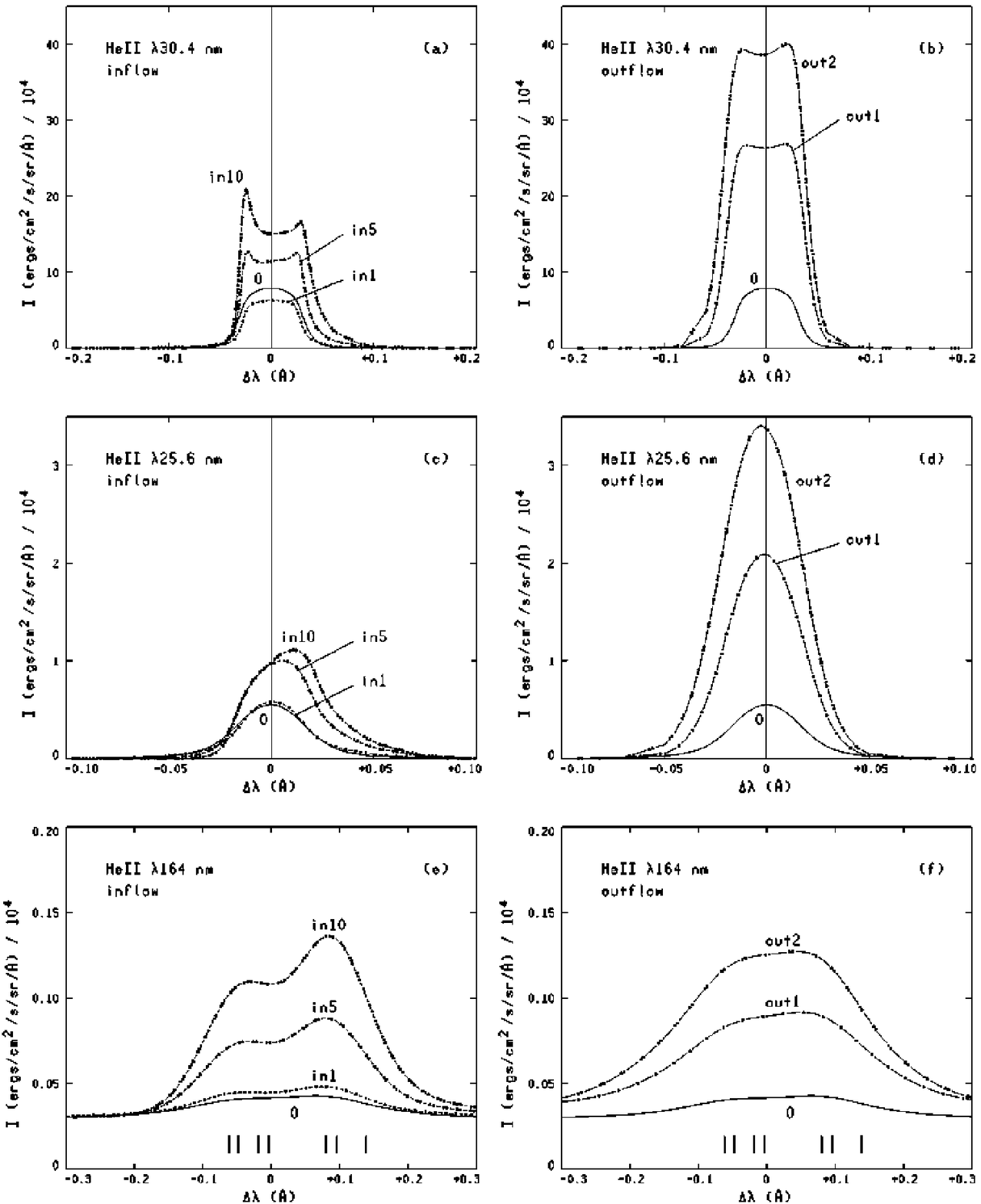}
\caption{ The calculated disk-center \ion{He}{2} line profiles
for the six energy-balance models. }
\end{figure}

\end{document}